\RequirePackage{fix-cm}
\documentclass[smallextended]{svjour3}       
\smartqed  
\usepackage{graphicx}
\usepackage{natbib}
\usepackage{amsmath}
\usepackage{xcolor}
\usepackage{siunitx} 
\newcommand\pyratbay{\textsc{Pyrat Bay~}}

\newcommand{\micron}{\SI{}{\um}}
\usepackage{xcolor}

\begin{document}

\title{A retrieval challenge exercise for the Ariel mission
}


\author{Joanna K. Barstow         \and
        Quentin Changeat \and Katy L. Chubb \and Patricio E. Cubillos \and Billy Edwards \and Ryan J. MacDonald \and Michiel Min \and Ingo P. Waldmann}

\authorrunning{Barstow, Changeat, Chubb, Cubillos, Edwards, MacDonald, Min, Waldmann} 

\institute{J. K. Barstow \at School of Physical Sciences, The Open University, Walton Hall, Milton Keynes, MK7 6AA, UK \\
\email{jo.barstow@open.ac.uk}             \\
   \and
Q. Changeat\at Department of Physics and Astronomy, University College London, Gower Street, London, WC1E 6BT, UK \\
\and K. L. Chubb\at SRON, Sorbonnelaan 2, 3584 CA, Utrecht, Netherlands\\
\and P. Cubillos\at Space Research Institute, Austrian Academy of Science, Schmiedlstrasse 6,A- 8042 Graz, Austria \\ \and
B. Edwards \at Department of Physics and Astronomy, University College London, Gower Street, London, WC1E 6BT, UK \\
\and R. J. MacDonald\at Department of Astronomy, Cornell University, 404 Space Sciences Building, Ithaca, New York 14853, USA \\
\and M. Min\at SRON, Sorbonnelaan 2,  3584 CA, Utrecht, Netherlands \\
\and I. P. Waldmann\at Department of Physics and Astronomy, University College London, Gower Street, London, WC1E 6BT, UK }

\date{Received: date / Accepted: date}

\maketitle

\begin{abstract}
The Ariel mission, due to launch in 2029, will obtain spectroscopic information for 1000 exoplanets, providing an unprecedented opportunity for comparative exoplanetology. Retrieval codes - parameteric atmospheric models coupled with an inversion algorithm - represent the tool of choice for interpreting Ariel data. Ensuring that reliable and consistent results can be produced by these tools is a critical preparatory step for the mission. Here, we present the results of a retrieval challenge. We use five different exoplanet retrieval codes to analyse the same synthetic datasets, and test a) the ability of each to recover the correct input solution and b) the consistency of the results. We find that generally there is very good agreement between the five codes, and in the majority of cases the correct solutions are recovered. This demonstrates the reproducibility of retrievals for transit spectra of exoplanets, even when codes are not previously benchmarked against each other. 

\end{abstract}

\section{Introduction}
In recent years, our knowledge of exoplanet atmospheres has been increasing rapidly.  Recent highlights have included the detection of hazes/clouds in most exoplanets \citep{sing16, barstow17, tsiaras18, pinhas19}, the presence of ionised metals in the atmosphere of an ultrahot Jupiter \citep{hoeijmakers18}, and the discovery of water vapour in the atmosphere of a small, temperate planet \citep{benneke19,tsiaras19}. Studies of their 3-dimensional structures have also highlighted the complexity of these worlds \citep{stevenson14b,arcangeli2019,demory2016}. In addition, atmospheric characterisation of directly imaged planets is providing interesting results \citep[see e.g.][]{2020A&A...640A.131M}.

The majority of planets for which we have obtained detailed atmospheric information transit their parent stars. Their atmospheres can be observed either during transit, when starlight passes through the limb of the atmosphere, or at eclipse, when a difference measurement between fluxes just outside of and during the eclipse reveals reflected light (in the optical) and thermal emission from the planet itself (in the infrared). For the most favourable targets, reflection or emission can be measured as a function of phase, providing a map of planetary conditions. These observations can be made using both space- and ground-based facilities. 

Typically, transit, eclipse and phase curve spectra are analysed using so-called `retrieval' modelling frameworks \citep{irwin08,madhusudhan2009,line13a,waldmann2015a,waldmann2015b,benneke2015,harrington16,lavie17,macdonald17,cubillos2018,gandhi2018,ormel19,Mollire19,kitzmann2019,  zhang2019,alrefaie19, min20}. Retrieval models incorporate a simplified, parameterised radiative transfer model, usually one-dimensional, and an algorithm to explore the parameter space and recover the model solution that provides the best fit to the data. Generally, these models involve minimal physical assumptions, instead allowing the atmospheric parameters to vary freely; as a consequence, this technique is a data-driven approach to spectral analysis. This is particularly advantageous for exoplanets, which often boast extremes of temperature and irradiation that stretch our understanding of atmospheric physics.

Ariel, \citep{tinetti16,pascale18,tinetti18}, a European Space Agency mission currently expected to launch in 2029, will perform the first census of transiting exoplanet atmospheres. Retrieval algorithms therefore have a critical role to play in support of mission. 

In this paper, we present a retrieval challenge conducted by the Spectral Retrievals Working Group for the Ariel Science Team. We use simulated Ariel observations to test our ability to recover a range of properties of transiting exoplanet atmospheres, using five independent retrieval frameworks that are currently used in the literature. We describe the basic properties of each model in Section~\ref{codes}; how the challenge was set up in Section~\ref{setup}; and the results in Section~\ref{results}.

\section{Retrieval Codes}
\label{codes}
The main details of the five retrieval schemes used in this analysis are briefly summarised here. For detailed information, we recommend referring to the journal articles for each. 

Each retrieval code conforms to the usual basic structure of a simple, parameterised radiative transfer model, coupled to an algorithm that samples the model parameters from a pre-defined prior distribution and converges towards the most likely solution. The versions of each model used in this work are 1D, and all contain the same model parameters, which are described in Section~\ref{setup}. All models except \pyratbay use a Nested Sampling approach for convergence, whilst \pyratbay uses an MCMC sampler. 

\subsection{ARCiS}
The ARtful modelling Code for exoplanet Science (ARCiS) is a forward modelling and Bayesian retrieval code designed to include physical and chemical atmospheric processes \citep{min20}. The structure of the atmosphere can be defined through classical parameterisations or computed self consistently using various approximations. For the physical and chemical processes, computationally efficient methods are included that are parameterised where our physical knowledge is lacking. For a more detailed description of the modelling philosophy we refer to \cite{min20} and \cite{2019A&A...622A.121O}. The radiative transfer is computed using correlated-k sampling of the molecular opacities. Many molecules are included, where available from the ExoMol database \citep{tennyson2016,chubb20}. Clouds can be included either parameterised or using the cloud formation concept from Ormel \& Min (2019). The cloud opacities are computed from either Mie theory or using a model for irregularly shaped particles \citep{2005A&A...432..909M}. Efficient isotropic multiple scattering calculations can be included with a correction factor for anisotropic scattering. Full anisotropic scattering can be performed as well using a Monte Carlo scattering method. Processes that can be computed include, chemical equilibrium and disequilibrium (Kawashima et al. in prep) and cloud and haze formation. Also the pressure temperature structure can be computed from radiative equilibrium with the stellar irradiation. The retrieval can be done using either optimal estimation or Multinest Bayesian sampling.

\subsection{NEMESIS}
NEMESIS is a retrieval scheme that works with both optimal estimation and nested sampling approaches. It incorporates a fast correlated-k radiative transfer model, where the correlated-k approximation is a way of pre-tabulating gas absorption coefficients within a wavelength interval, relying on the assumption that the strongest lines at one level in the atmosphere are correlated with the strongest lines at other levels. For further details see \cite{irwin08}, \cite{krissansen-totton18}, \cite{goodyyung} and \cite{lacis91}. NEMESIS was originally developed for analysis of Solar System planets (e.g. \citealt{fletcher09,tsang10}) and has subsequently been extended to exoplanets (e.g. \citealt{lee12,barstow14}).

Line data are sourced primarily from the ExoMol project and provided in appropriate format for each model by \cite{chubb20}. H$_2$O is from \cite{pokazatel}, CO$_2$ from \cite{huang17}, CO from \cite{li15}, CH$_4$ from \cite{yurchenko17} and TiO from \cite{mckemmish19}. 

\subsection{Pyrat Bay}

The Python Radiative Transfer in a Bayesian framework ({\pyratbay}, \citealt{CubillosBlecic2021mnrasPyratBay}), is a modular open-source code to model exoplanet spectra and retrieve the planet's atmospheric properties.
The atmospheric models consist of parameterized 1D profiles of the temperature, composition, and altitude (in hydrostatic-equilibrium) as a function of pressure.  For transmission geometry, {\pyratbay} solves the radiative-transfer equation under the plane-parallel approximation, sampling the opacities at a constant resolving power spectrum, over the wavelengths considered here.

{\pyratbay} considers opacities from the main sources expected for
exoplanets at these wavelengths: molecular line
transitions from from HITRAN or ExoMol \cite{RothmanEtal2010jqsrtHITEMP, GordonEtal2017jqsrtHITRAN2016, TennysonYurchenko2018atomsExomol}, collision-induced absorption from Borysow or HITRAN \citep{BorysowFrommhold1989apjH2HeOvertones, BorysowEtal1989apjH2HeRVRT,
  BorysowEtal2001jqsrtH2H2highT, Borysow2002jqsrtH2H2lowT,karman2019}, 
resonant Na and K opacity models \citep{BurrowsEtal2000apjBDspectra},
  Rayleigh scattering for H, He, and H$_{2}$ \citep{Kurucz1970saorsAtlas,
  LecavelierDesEtangsEtal2008aaRayleighHD189},  and several cloud
models, from a simple gray cloud deck to complex Mie-scattering \citep{ToonAckerman1981apoptScattering} models in thermal stability
\citep{ackerman01} or microphysical parameterization (Blecic et al., in prep.).  {\pyratbay} handles the billion-sized line lists by compressing them with the
\textsc{repack} package \citep{Cubillos2017apjCompress}, to extract only the
dominating line transitions.

The code explores the parameter space via a differential-evolution
MCMC sampler implemented in
\citep{CubillosEtal2017apjRednoise}, checking on the
Gelman--Rubin statistics for convergence \citep{GelmanRubin1992}.

\subsection{TauREx}
TauREx (Tau Retrieval for Exoplanets) is a fully Bayesian radiative  transfer  and  retrieval framework \citep{waldmann2015a,waldmann2015b,alrefaie19}. TauREx can be used with the line-by-line cross sections from the Exomol project \citep{tennyson2016} and HITEMP \citep{rothman14} and HITRAN \citep{gordon16}.  TauREx can be used to model both transmission and thermal emission. We also included absorptions from Rayleigh scattering and CIA for the couples H$_2$-H$_2$ and H$_2$-He \citep{Borysow2001,borysow02, Rothman2013}.  The public version of TauREx is able to retrieve chemical composition  of  exoplanets  by  assuming  constant abundances with altitude,  parametric 2-layer variations \citep{changeat19}, or equilibrium chemistry \citep{venot01}. In the new version, TauREx\,3 is particularly flexible, allowing users to redefine any part of the code with their own custom modules. To perform the retrieval, TauREx can use multiple sampling techniques. Here we use the nested sampling retrieval algorithm Multinest \citep{feroz09} in its python implementation PyMultinest \citep{buchner14}.

\subsection{POSEIDON}
POSEIDON is a nested sampling retrieval code for exoplanet transmission spectra \citep{macdonald17}. Radiative transfer is computed via the sampling of high spectral resolution ($R \sim 10^6$) cross sections onto intermediate resolution wavelength grids (typically $100 \times$ higher than the resolution of the observations being retrieved), producing a close representation of line-by-line radiative transfer. The atmospheric temperature structure can be parameterised either via a 6-parameter function \citep{madhusudhan2009} or an isotherm. Inhomogenous `patchy' clouds and hazes are included, allowing cloud fractions to be retrieved. Over 50 chemical species are currently supported as retrievable parameters, with molecular line data largely sourced from ExoMol \citep{tennyson2016}, atomic data from VALD3 \citep{ryabchikova2015}, and continuum data from HITRAN \citep{karman2019}. 

\section{Retrieval Challenge} 
\label{challenge}
\subsection{Setup}
\label{setup}
Here, we present the results of a multi-code retrieval challenge, conducted using synthetic spectra generated by TauREx with retrievals by NEMESIS, ARCiS, Pyrat Bay, TauREx and POSEIDON. 
Four spectra, representing respectively a clear hot Jupiter (with system parameters similar to e.g. HAT-P-30b), a cloudy hot Jupiter (with system parameters similar to e.g. XO-2Nb), a clear warm Neptune (with system parameters similar to e.g. GJ436b) and a cloudy warm super-Earth (with system parameters similar to e.g. GJ1214b) were provided, with known atmospheric inputs. Appropriate noise for Ariel was generated using the radiometric model ArielRad \citep{mugnai19}, and added as a error envelope to each synthetic observation. This allowed each user to test, benchmark and modify their retrieval procedure. A further four spectra representing similar planets, but without known atmospheric properties, were also provided such that a blind retrieval tests could be conducted with NEMESIS, ARCiS, Pyrat Bay and POSEIDON. All retrievals with TauREx were non-blind, as this model was used to generate the synthetic spectra. This also meant that for TauREx retrievals only included parameters known to be necessary for each case. For the other four codes, H$_2$O, CO$_2$, CO, CH$_4$, TiO and cloud are included in all retrievals, regardless of whether they were included in the input to the synthetic spectrum. Note that the only fundamental difference between the blind and non-blind retrievals is the prior knowledge of the person executing the retrievals and thus the inclusion of certain parameters.

The setup of all atmospheric forward and retrieval models consists of an isothermal atmosphere with constant abundances of the given molecular species. The atmospheres are assumed to be in hydrostatic equilibrium with the mean molecular weight computed using the molecular abundances plus 85\% H$_2$ and 15\% He. The cloud is modeled using a grey, infinite opacity cloud deck at all pressure above a certain pressure level. The spectra were simulated to represent Ariel Tier 2 data \citep[for details we refer to table 1 in][]{2020AJ....160...80C}.

\begin{table*}[t] 
\centering 
\begin{tabular}{c c c c c c c c c} 
\hline 
Planet & R$_S$ (R$_\odot$) & M$_P$ (M$_J$) & R$_P$ (R$_J$) & H$_2$O & CH$_4$ & CO & Temp (K) & P$_{cloud}$ (bar)\\ 
\hline 
1 & 1.2 & 0.7 & 1.35 & 7.0$\times$10$^{-6}$ & 0.0 & 4.5$\times$10$^{-4}$ & 1900 & None \\ 
1B & \textbf{1.2} & \textbf{0.73} & 1.4 & 1.0$\times$10$^{-5}$ & 0.0 & 1.0$\times$10$^{-3}$ & 1600 & None \\ 
2 & 1.2 & 0.7 & 1.35 & 7.0$\times$10$^{-6}$ & 0.0 & 4.5$\times$10$^{-4}$ & 1900 & 1.94$\times$10$^{-2}$\\ 
2B & \textbf{1.1} & \textbf{0.7} & 1 & 4.0$\times$10$^{-5}$ & 0.0 & 7.0$\times$10$^{-4}$ & 1500 & 8.74$\times$10$^{-5}$ \\ 
3 & 0.5 & 0.06 & 0.3 & 2.0$\times$10$^{-6}$ & 5.0$\times$10$^{-5}$  & 0.0 & 700 & None \\ 
3B & \textbf{0.45} & \textbf{0.07} & 0.35 & 1.0$\times$10$^{-5}$ & 1.0$\times$10$^{-5}$ & 0.0 & 750 & None \\
4 & 0.3 & 0.01 & 0.11 & 2.0$\times$10$^{-4}$ & 1.0$\times$10$^{-4}$ & 0.0 & 1000 & 2.91$\times$10$^{-2}$ \\
4B & \textbf{0.45} & \textbf{0.01} & 0.15 & 1.0$\times$10$^{-6}$ & 1.0$\times$10$^{-5}$ & 0.0 & 700 & 4.52$\times$10$^{-2}$ \\
\hline 
\end{tabular}
\smallskip 
\caption{Input model parameters for each planet case. Numbers with suffix B indicate the blind test version. Parameter values shown in bold text for the blind versions indicate the known parameters. All parameters are known for the non-blind cases.} 
\label{properties} 
\end{table*}

The bulk planet properties for each case are listed in Table~\ref{properties}. For the blind retrievals, only the planet mass and stellar radius are assumed to be known. 

\subsection{Results}
\label{results}
\begin{figure*}[!h]
\centering
\includegraphics[width=0.9\textwidth]{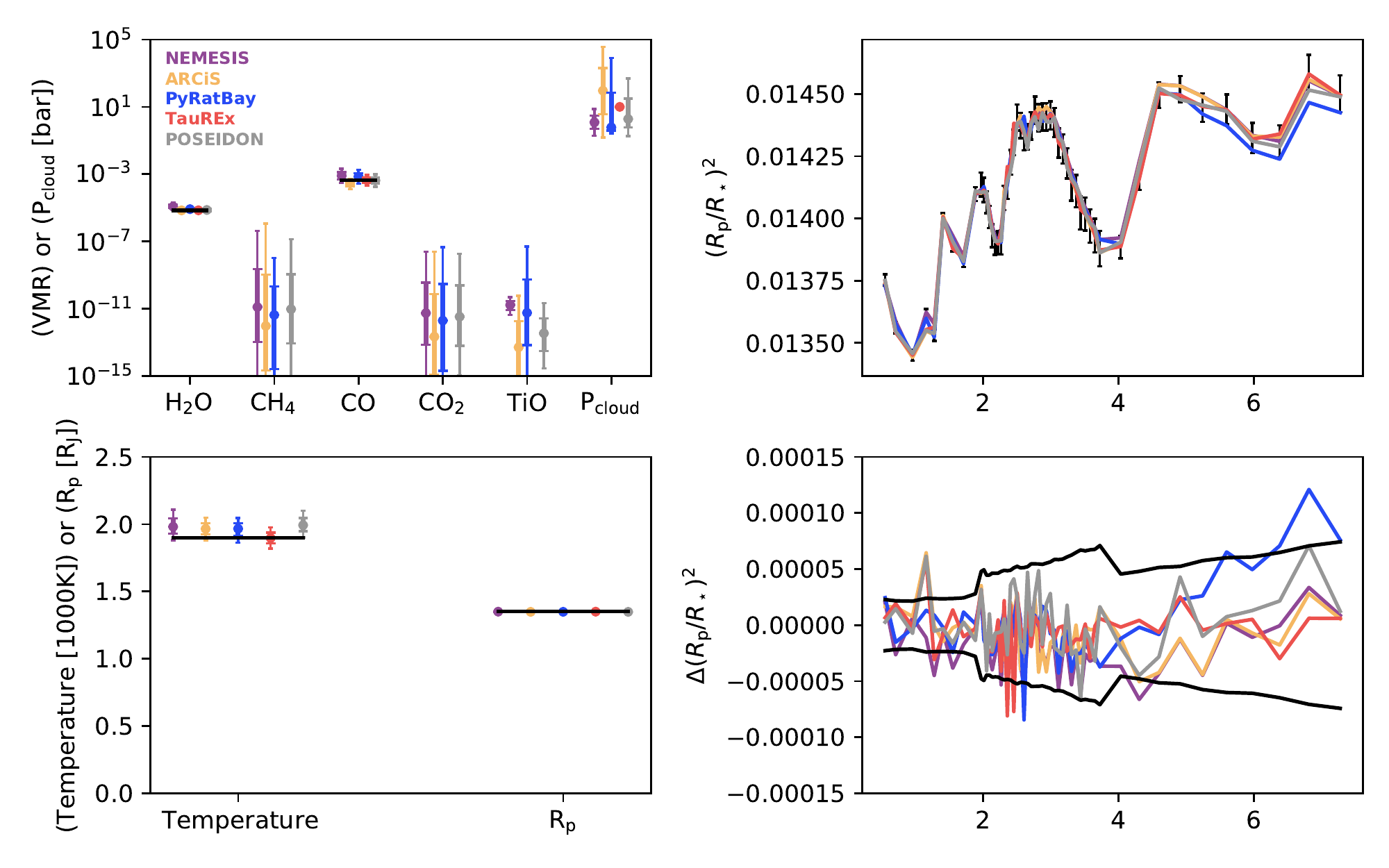} 
\caption{Retrieval results and spectral fits for Planet 1. The colours represent the different retrievals used. Black lines on the parameter plots (left hand panels) indicate the input values for retrieved quantities. Where black lines and TauREx retrieved values are absent, the gas/cloud was not included in the input model. Thick/thin error bars indicate the 1/2-$\sigma$ limits respectively. The black points in the top right panel indicate the input spectrum with error bars. The difference spectra (bottom right) show input - model for each retrieval, with the black lines indicating the error envelope. The x-axis for the right hand plots shows wavelength in microns.}
\label{planet1}
\end{figure*}

\begin{figure*}[!h]
\centering
\includegraphics[width=0.9\textwidth]{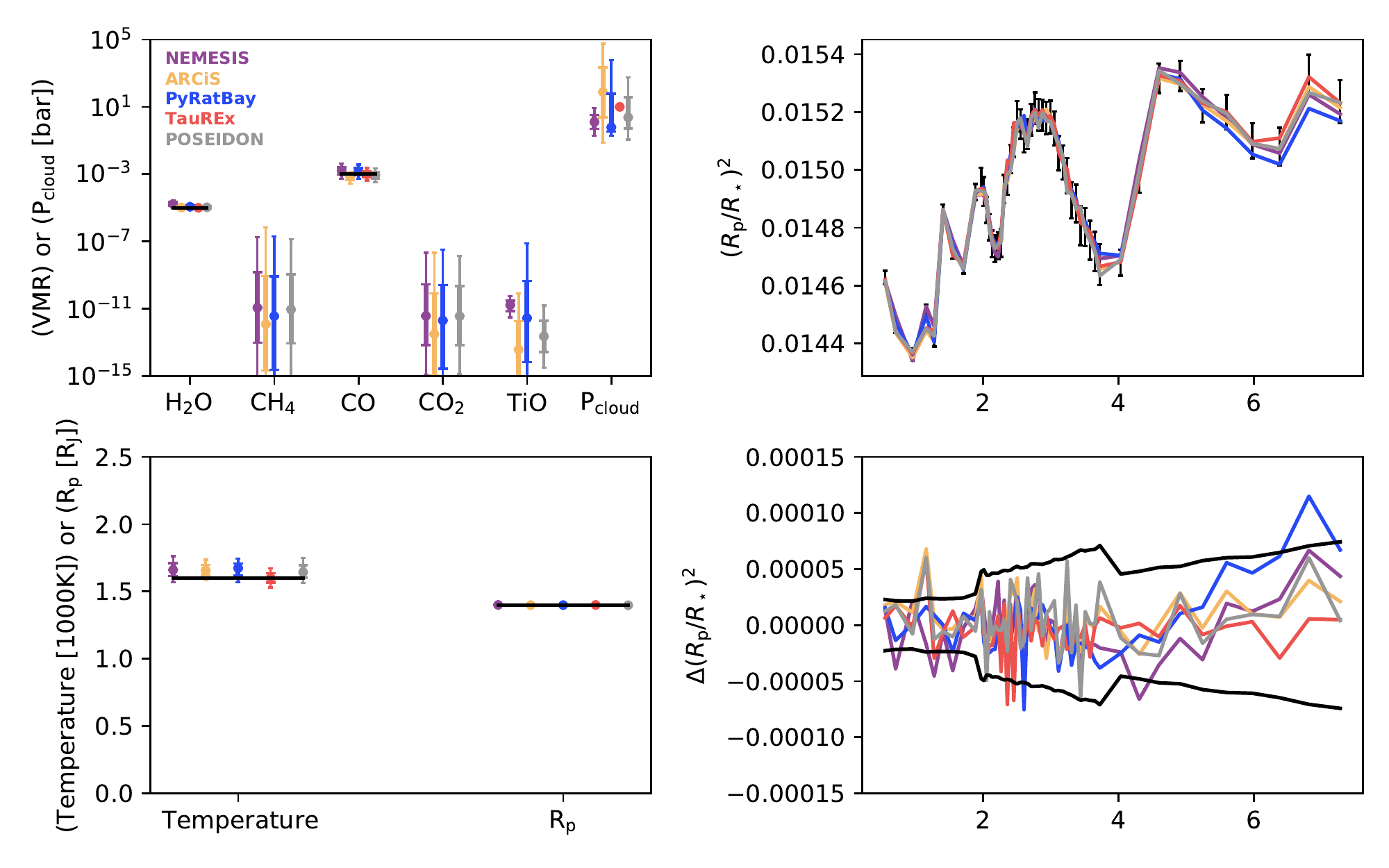} 
\caption{As Figure~\ref{planet1} but for Planet 1B.}
\label{planet1b}
\end{figure*}

\begin{figure*}[!h]
\centering
\includegraphics[width=0.9\textwidth]{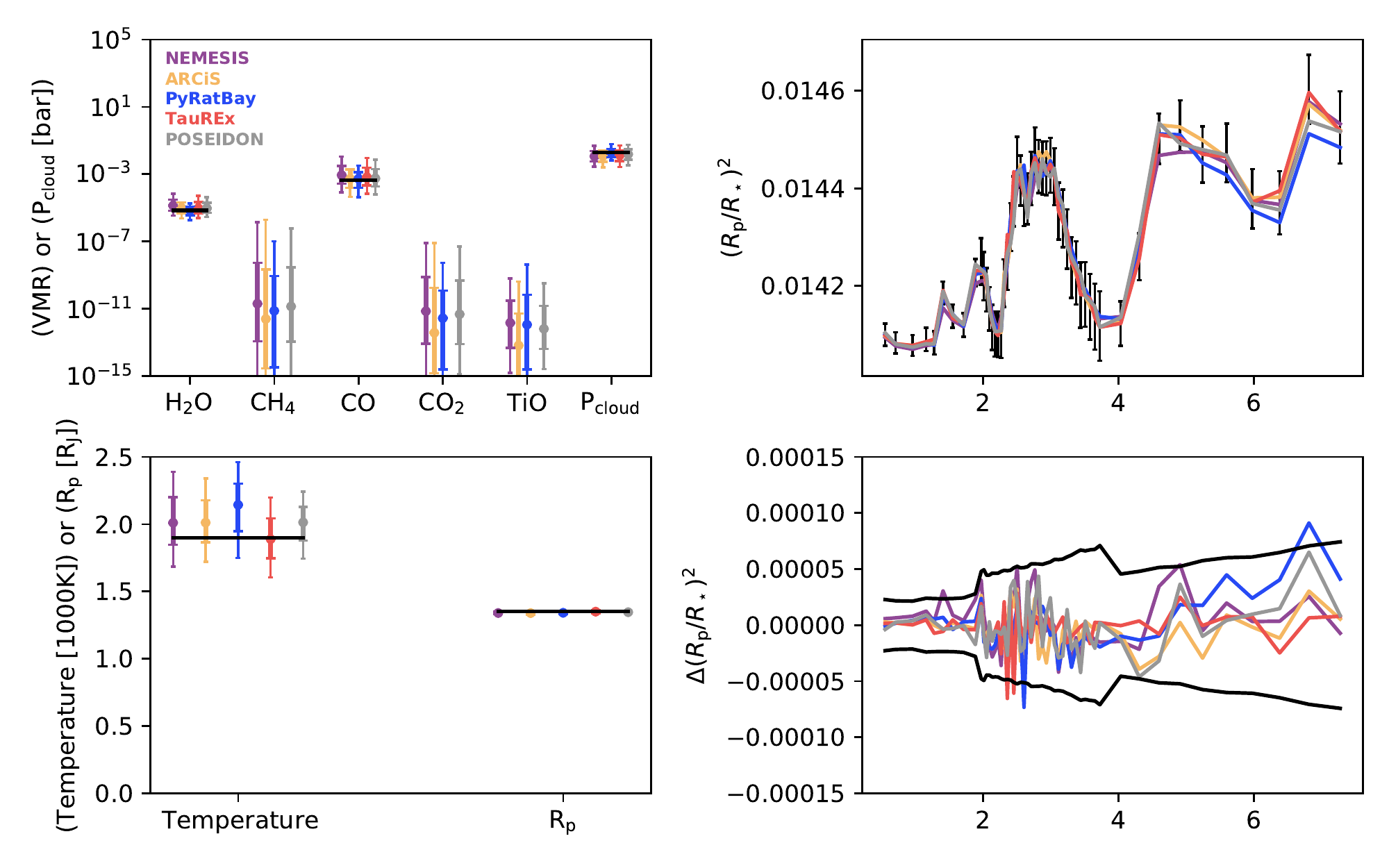} 
\caption{As Figure~\ref{planet1} but for Planet 2.}
\label{planet2}
\end{figure*}

\begin{figure*}[!h]
\centering
\includegraphics[width=0.9\textwidth]{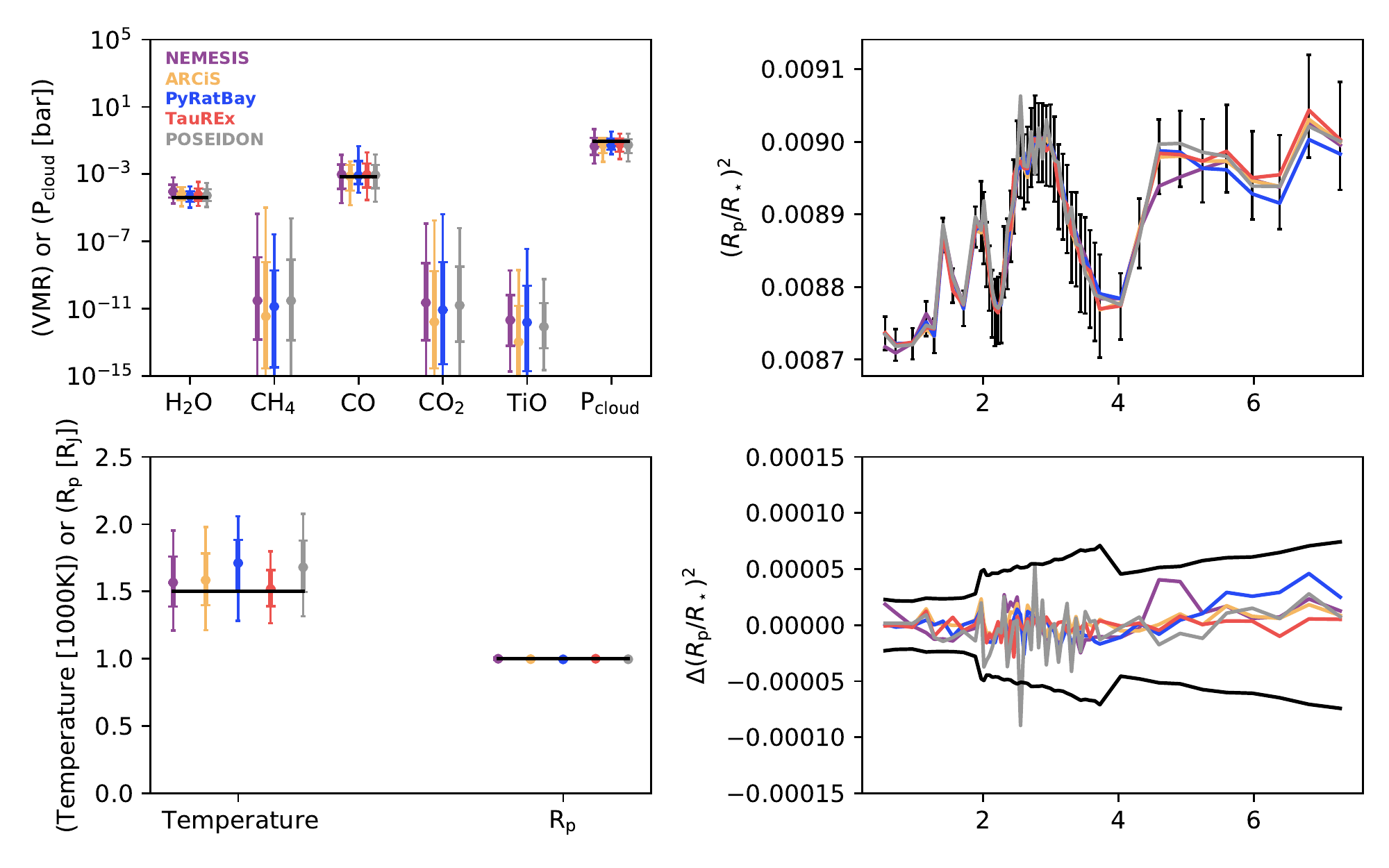} 
\caption{As Figure~\ref{planet1} but for Planet 2B.}
\label{planet2b}
\end{figure*}

\begin{figure*}[!h]
\centering
\includegraphics[width=0.9\textwidth]{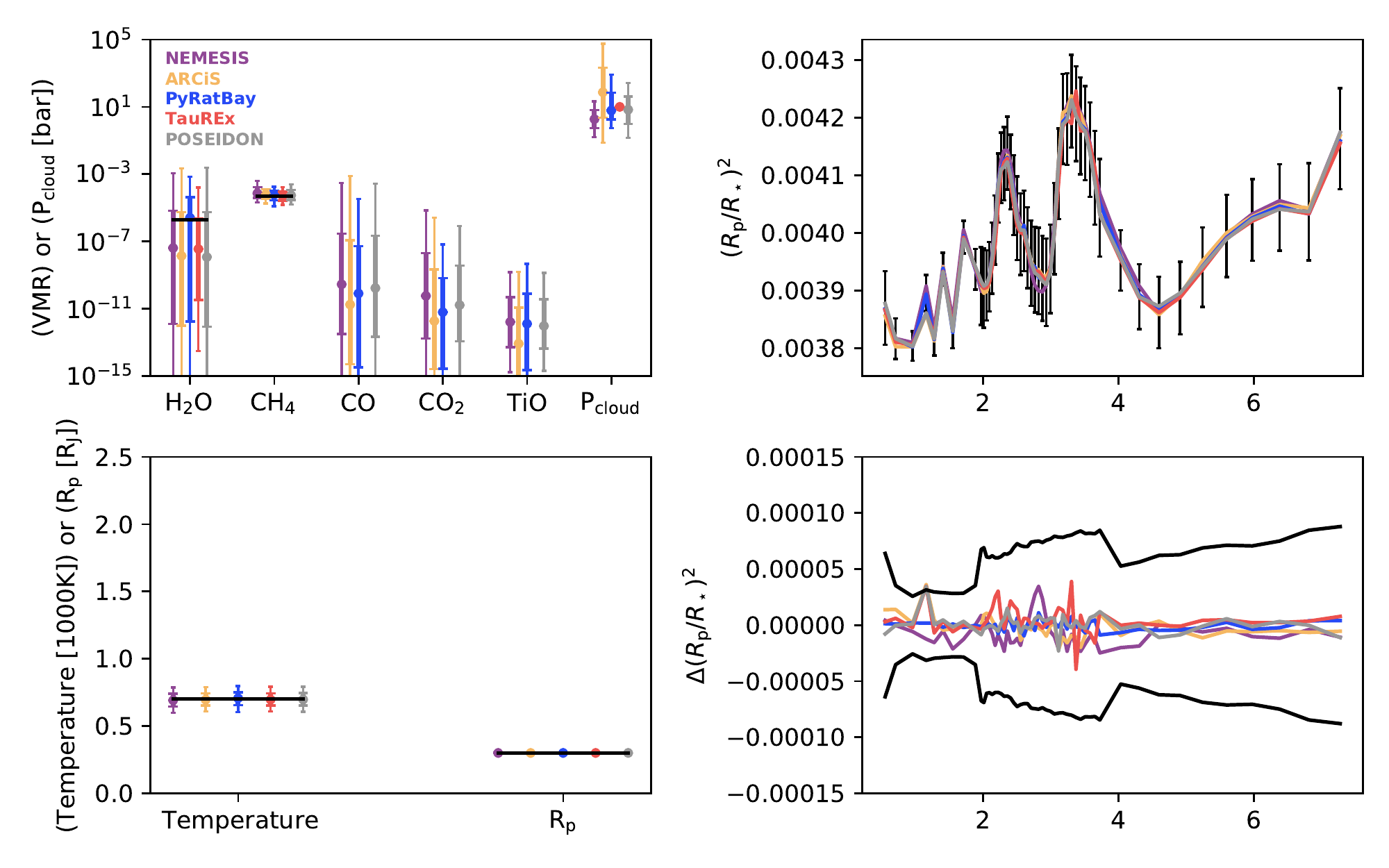} 
\caption{As Figure~\ref{planet1} but for Planet 3.}
\label{planet3}
\end{figure*}

\begin{figure*}[!h]
\centering
\includegraphics[width=0.9\textwidth]{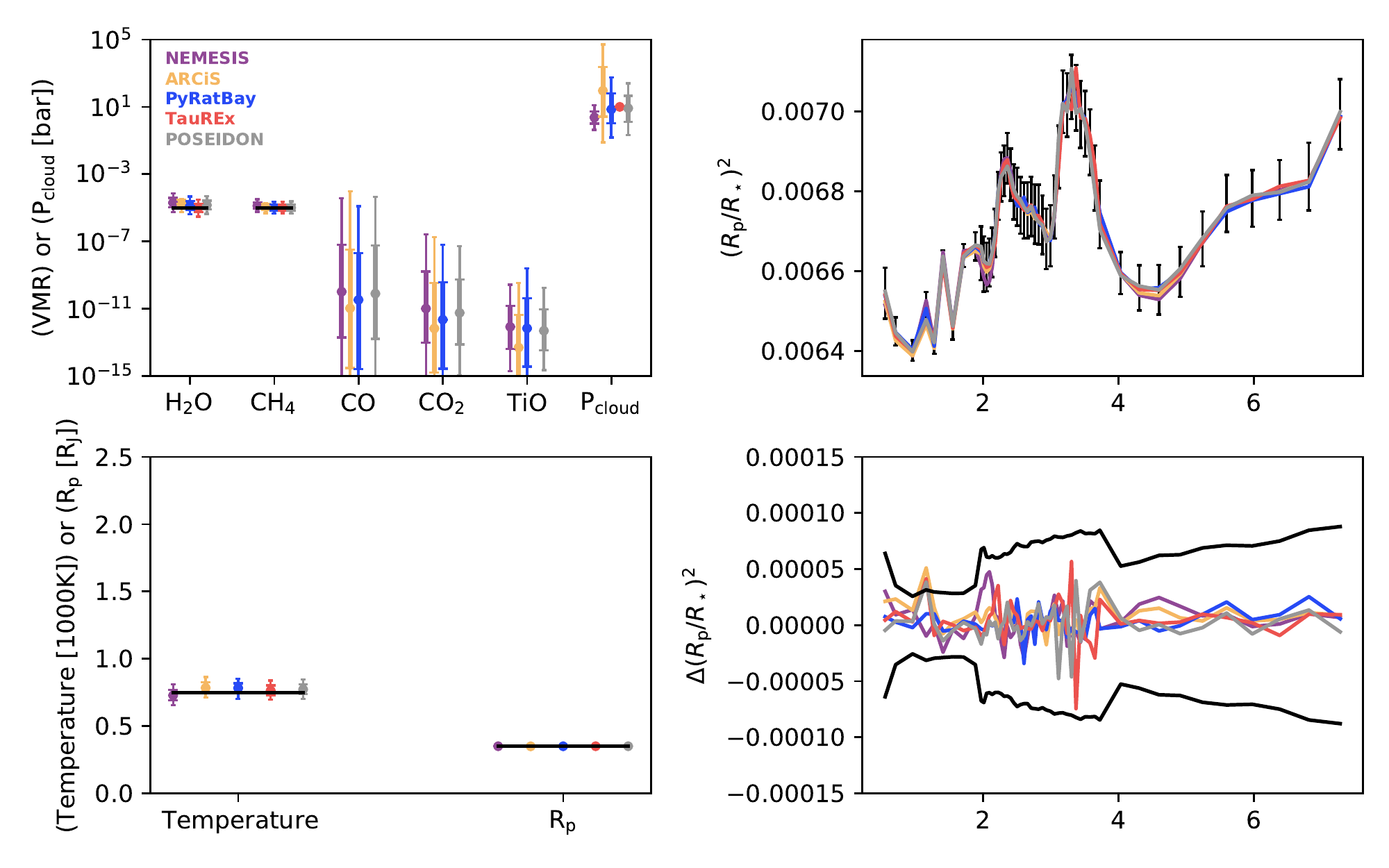} 
\caption{As Figure~\ref{planet1} but for Planet 3B.}
\label{planet3b}
\end{figure*}

\begin{figure*}[!h]
\centering
\includegraphics[width=0.9\textwidth]{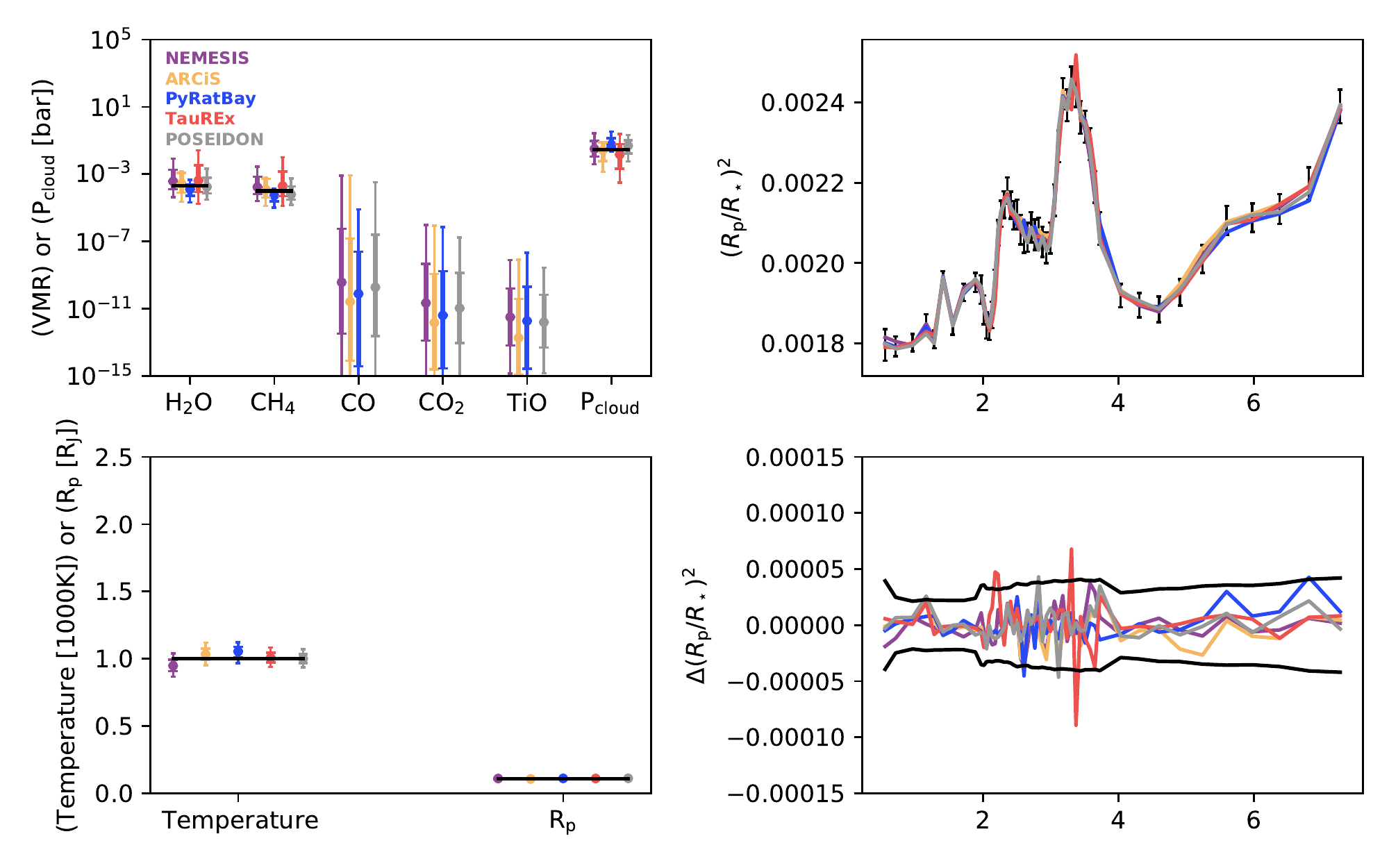} 
\caption{As Figure~\ref{planet1} but for Planet 4.}
\label{planet4}
\end{figure*}

\begin{figure*}[!h]
\centering
\includegraphics[width=0.9\textwidth]{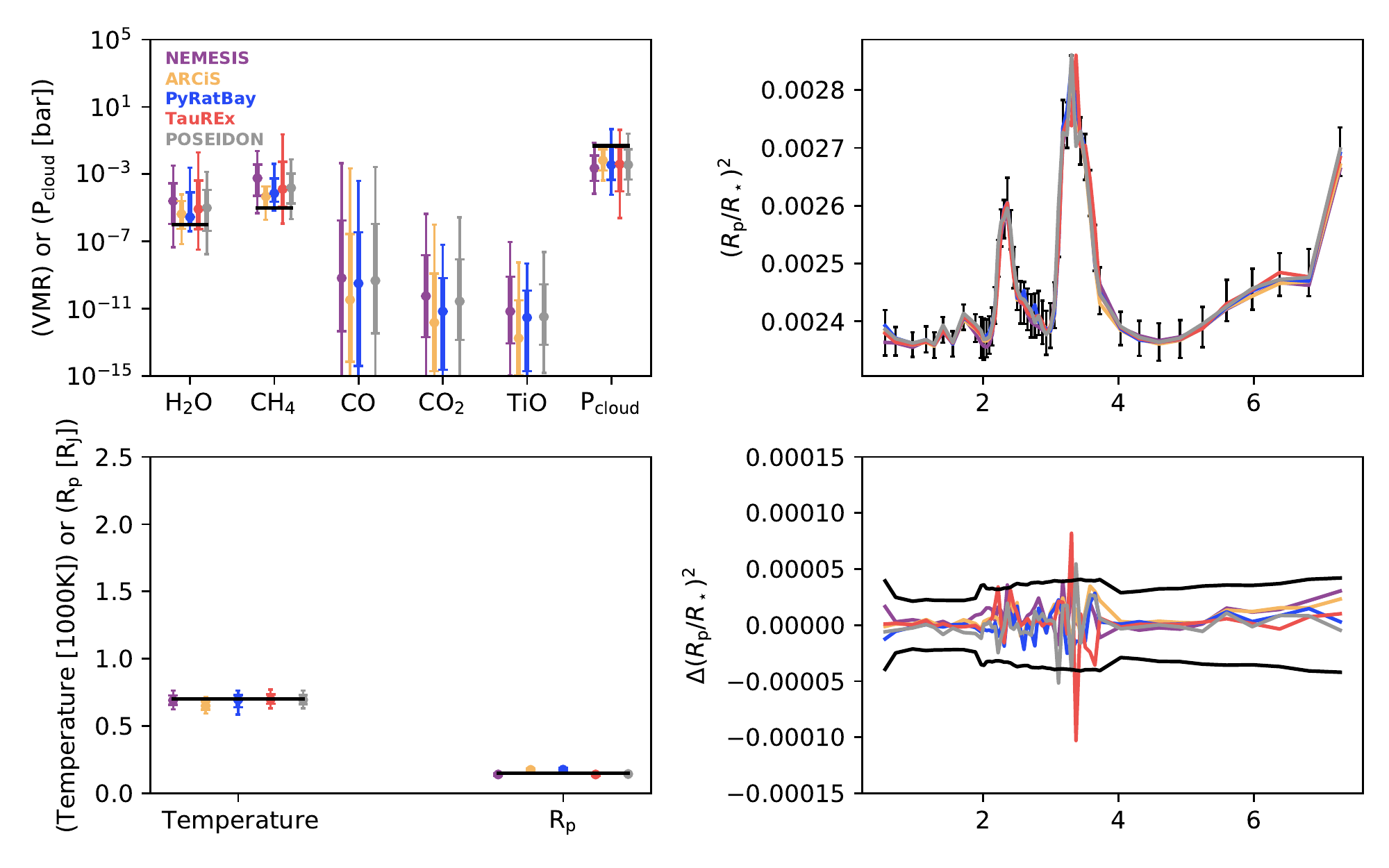} 
\caption{As Figure~\ref{planet1} but for Planet 4B.}
\label{planet4b}
\end{figure*}

The combined retrieval results from all five codes are compared with the input values, with 1$\sigma$ error bars, in Figures~\ref{planet1}---\ref{planet4b}. We also present example corner plots for each code, for Planet 2, in Figures~\ref{posterior_nem}---\ref{posterior_poseidon}; these provide a better indication of correlations and degeneracies between parameters. 

The results shown here display overall very good agreement between both spectra and retrieved parameters for all cases, both given and blind. In general, retrieved parameters are also correct to within 1$\sigma$ from the input value.

\subsubsection{Spectral fits}
The quality of the spectral fits is generally extremely good. The $\chi^2$ values for each model and planet are presented in Table~\ref{chisq}. 

There are specific spectral regions where discrepancies emerge for some models. For example, for \pyratbay there is generally a discrepancy at around 5.5\,\micron; \pyratbay underestimates the TauREx transit depth consistently here. Similarly, NEMESIS underestimates the transit depth at around 1.3\,\micron~in Figure~\ref{planet2}, but overestimates it in Figure~\ref{planet1}. These discrepancies across small wavelength regions are most likely to be related to different treatments of absorption line data. ARCiS and NEMESIS use k tables, whilst the other codes use cross sections. These different methods for tabulating and binning absorption data both introduce some error over the (much less efficient) line-by-line approach. Additionally, different cross-section or k table grids could also be a source of error between codes that use the same method. This is explored in more detail in Section~\ref{pyratbay_crosssection}.

\subsubsection{Retrieval results}
Retrieved results are correct to within 2$\sigma$ in all cases except for a few radius retrievals. Since the planet radius is the most precisely determined quantity, small offsets in the synthetic spectrum can produce large deviations in the measured radius with respect to the error bars. This does not however affect the accuracy of the other retrieved quantities. 

The one remaining discrepant result is in the temperature retrieval for Planet 3B, for the ARCiS and Pyrat Bay models. This spectrum has relatively large error bars, allowing more flexibility in the model fit than some of the other examples. In addition, the NEMESIS and TauREx codes have been substantially benchmarked against each other (\cite{barstow20}) so NEMESIS may be expected to more faithfully reproduce a spectrum generated using TauREx than other models that have been less extensively calibrated. POSEIDON retrieves the correct input result despite a lack of extensive prior benchmarking, which is testament to the generally high reliability of current spectral retrieval schemes. 

Where the retrieved gas is absent from the input model, each retrieval scheme with the exception of TauREx is effectively retrieving an upper limit on that gas. Generally, the error bars for these species are large, indicating little constraint. The exception to this is an apparently constrained TiO abundance for the clear hot Jupiter case from NEMESIS. This is likely to be a function of the fact that TiO features are not resolved by Ariel as they occur at short wavelengths, where the resolving power is low, so a small amount of TiO can be invoked to produce opacity at shorter wavelengths without there being confidence that TiO is actually the species concerned. Definitive detection of TiO with Ariel is likely to be a challenge as a result of this. 

Where there is no cloud in the input model, an upper limit for the cloud top pressure is retrieved. This upper limit corresponds to the pressure at which the atmosphere becomes opaque in transmission (so any cloud deck sitting beneath this level would be invisible). 

The generally very good agreement in both spectral fit and retrieved properties from five different retrieval models, and the accuracy of the retrieved solutions, demonstrates that a wealth of atmospheric information can be reliably recovered from Ariel spectra. 

\begin{table}
\centering
\begin{tabular}{c|c c c c c}
 Planet & NEMESIS & ARCiS & PyRat Bay & TauREx & POSEIDON\\
 \hline
 1  & 24.7 & 9.6 & 17.8 & 12.1 & 12.4 \\
 1B & 22.3 & 7.4 & 15.2 & 10.0 & 10.6 \\
 2 & 11.7 & 5.4 & 8.8 & 5.8 & 8.9 \\
 2B & 5.4 & 1.5 & 3.1 & 1.7 & 10.2 \\
 3 & 3.5 & 0.7 & 0.4 & 1.7 & 0.5 \\
 3B & 5.6 & 2.1 & 1.4 & 3.2 & 2.2 \\
 4 & 8.2 & 5.9 & 7.1 & 17.6 & 6.6 \\
 4B & 8.2 & 5.1 & 4.8 & 16.6 & 8.4 \\
 \hline
\end{tabular}
\caption{$\chi^2$ values for each model and planet fit. The spectrum has 52 data points and there are 8 free parameters in each case (fewer for TauREx), so the reduced $\chi^2$ values would are consistently below 1.}
\label{chisq}
\end{table}

\begin{table}
\centering
\begin{tabular}{c|c c c c c}
 Planet & NEMESIS & ARCiS & PyRat Bay & TauREx & POSEIDON\\
 \hline
 1	&	2.12	&	1.78	&	1.23	&	0.006	&	1.22	\\
1B	&	1.57	&	1.18	&	1.10	&	0.0007	&	0.94	\\
2	&	0.60	&	0.44	&	0.58	&	0.11	&	0.29	\\
2B	& 0.27	&	0.19	&	0.51	&	0.08	&	0.34	\\
3	&	0.14	&	0.05	&	0.03	&	0.16	&	0.36	\\
3B	&	0.47	& 0.82	&	0.29	&	0.04	&	0.19	\\
4	&	0.37	&	0.25	&	1.28	&	0.14	&	0.23	\\
4B	&	1.82	&	3.08	&	1.78	&	0.56	&	1.03	\\
 \hline
\end{tabular}
\caption{`Accuracy index' for each retrieval. This is based on the $\chi^2$ calculation for goodness of fit and described in more detail below. Lower values indicate more accurate retrievals.}
\label{accuracy}
\end{table}

\section{Discussion}
Here, we discuss the retrieved results for each model in more detail, and investigate further some discrepancies that emerge. It is necessary to note here that differences in output spectra and in model parameters of the level that we see here are likely to be equivalent to discrepancies between models and real data/truth; in a real-observation scenario, these differences could be due to uncorrected instrument systematics or astrophysical noise (e.g. stellar activity that is not accounted for), as well as incompleteness of the model itself \citep{barstow20}. 

\subsection{Retrieval accuracy}
In Table~\ref{accuracy} we present indices for the accuracy of each retrieval, in terms of its ability to correctly identify the value of the input parameters to within 1$\sigma$. This is defined in a similar way to the $\chi^2$ value. We only consider the parameters for each case that are constrained by the retrieval, so for example where a gas is not included in the input model the accuracy for the null detection is not calculated.

Our metric is given by:
\begin{equation}
  a_{\mathrm{index}}=\dfrac{\sum_{0}^{n}(x_{\mathrm{ret,n}}-x_{\mathrm{input,n}})^2/\sigma_{\mathrm{n}}^2}{n}  
\end{equation}

where $x_{\mathrm{input}}$ and $x_{\mathrm{ret}}$ are the input and retrieved values for each parameter, $\sigma$ is the (average) error on the parameter, and $n$ is the number of constrained parameters. We took logs for the volume mixing ratios and cloud pressure for this calculation. 

As shown in Table~\ref{accuracy}, retrieval accuracy varies between codes and planets, but in general retrievals for different codes are consistent with each other for a given planet. TauREx generally has an extremely low accuracy index, which is expected, since TauREx was used to generate the input model. 

The highest accuracy indices, indicating relatively poorer retrievals, are found for planets 1, 1B and 4B. Planets 1 and 1B also have relatively high $\chi^2$ values, indicating that these have some of the poorer spectral fits. By contrast, Planet 4B has a reasonable quality of fit, but since this is a cloudy planet this adds to the complexity of the retrieval. 

Examining the retrieved values for planets 1 and 1B reveals that the temperatures are slightly overestimated for both planets outside of the 1$\sigma$ error bars, except for the TauREx retrieval. This is likely to have resulted in the poorer accuracy values. For Planet 4B, the gas abundances are slightly overestimated for some codes, and the cloud top pressure is correspondingly underestimated, indicating that it is degeneracy between cloud pressure and gas abundances that is responsible for the relative lack of accuracy in this case.

For all other planets, the accuracy index is below 1 regardless of the retrieval model used (except Planet 4 for \pyratbay), indicating that all retrieved values are recovered correctly to within 1$\sigma$. The \pyratbay retrieval for Planet 4 suffers from similar cloud top pressure/gas abundance degeneracy to that seen more widely for Planet 4B. 

Despite this, the retrievals are accurate to within 2 $\sigma$ across all models, for all planets, and in most cases the molecular abundances are recovered correctly to within 1 $\sigma$. Considering that measurement of atmospheric composition is a key goal for Ariel, this finding provides confidence in the ability of the mission to deliver on its objectives. 

\subsection{Retrieval correlations}
For ease of comparison, we have so far shown simply the median, and 1- and 2-$\sigma$ values for each retrieved property. This of course doesn't show any correlations present between parameters, or fully capture the shape of the retrieved probability distribution. 

\begin{figure*}[!h]
\centering
\includegraphics[width=1.0\textwidth]{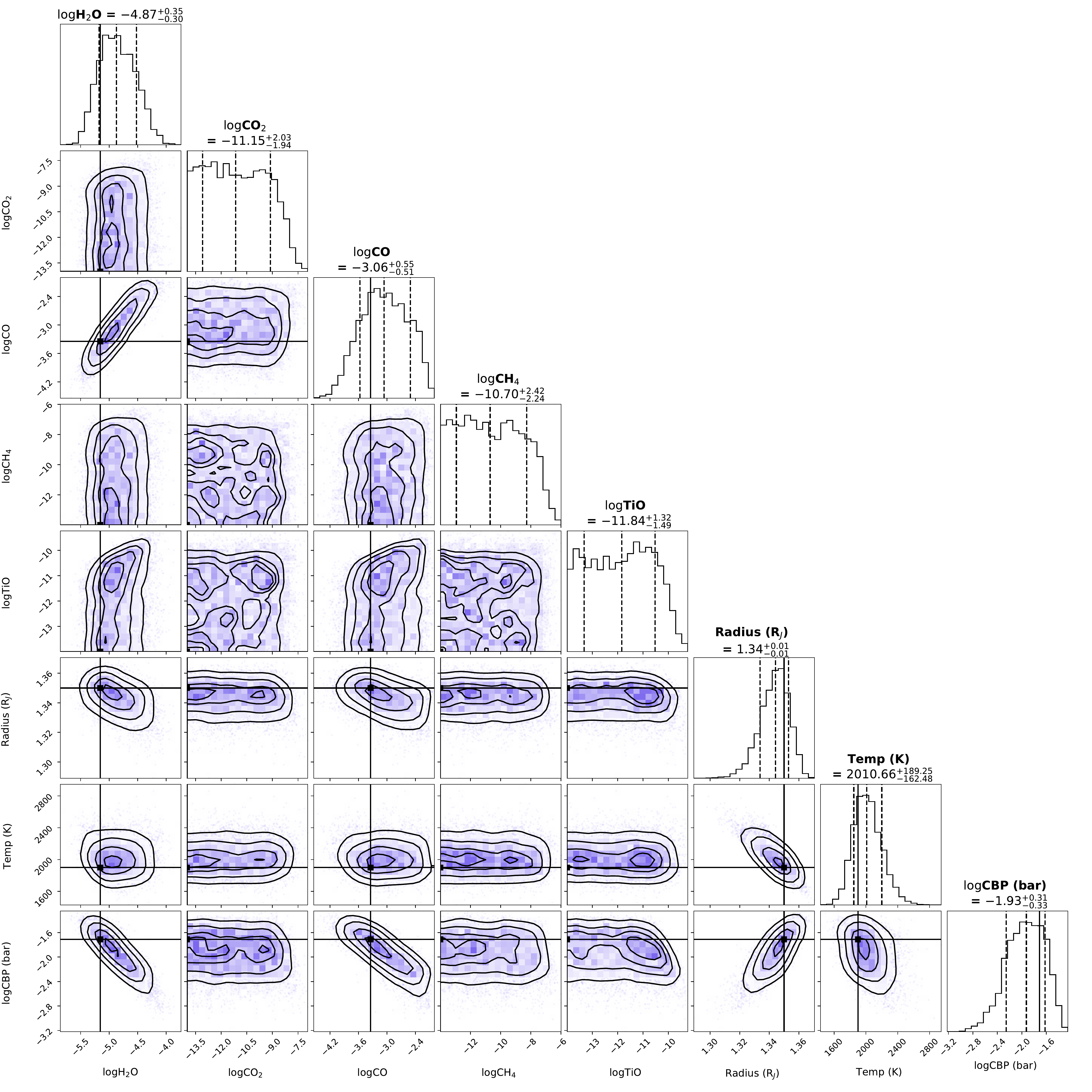}
\caption{This figure shows the retrieved posterior probability distributions for NEMESIS on Planet 2. Input values are indicated by black lines. Dashed lines show the median and +/-$\sigma$ values. In all cases, the true values fall within the +/-$\sigma$ range.}
\label{posterior_nem}
\end{figure*}

\begin{figure*}[!h]
\centering
\includegraphics[width=1.0\textwidth]{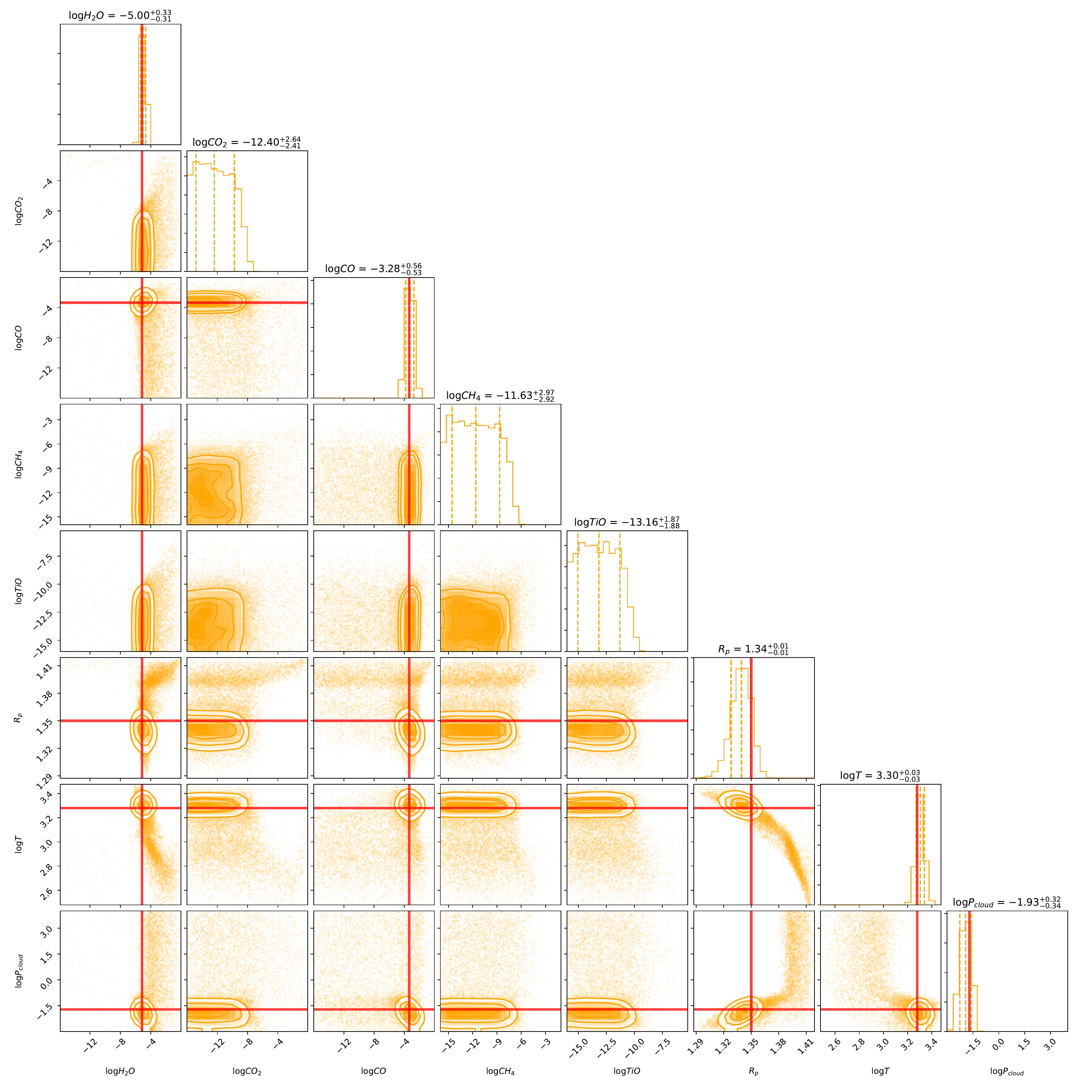}
\caption{As Figure \ref{posterior_nem} but for ARCiS. In this case, solid red lines indicate the input values and dashed lines the indicate the median retrieved values and the +/-$\sigma$ envelope.}
\label{posterior_arcis}
\end{figure*}

\begin{figure*}[!h]
\centering
\includegraphics[width=1.0\textwidth]{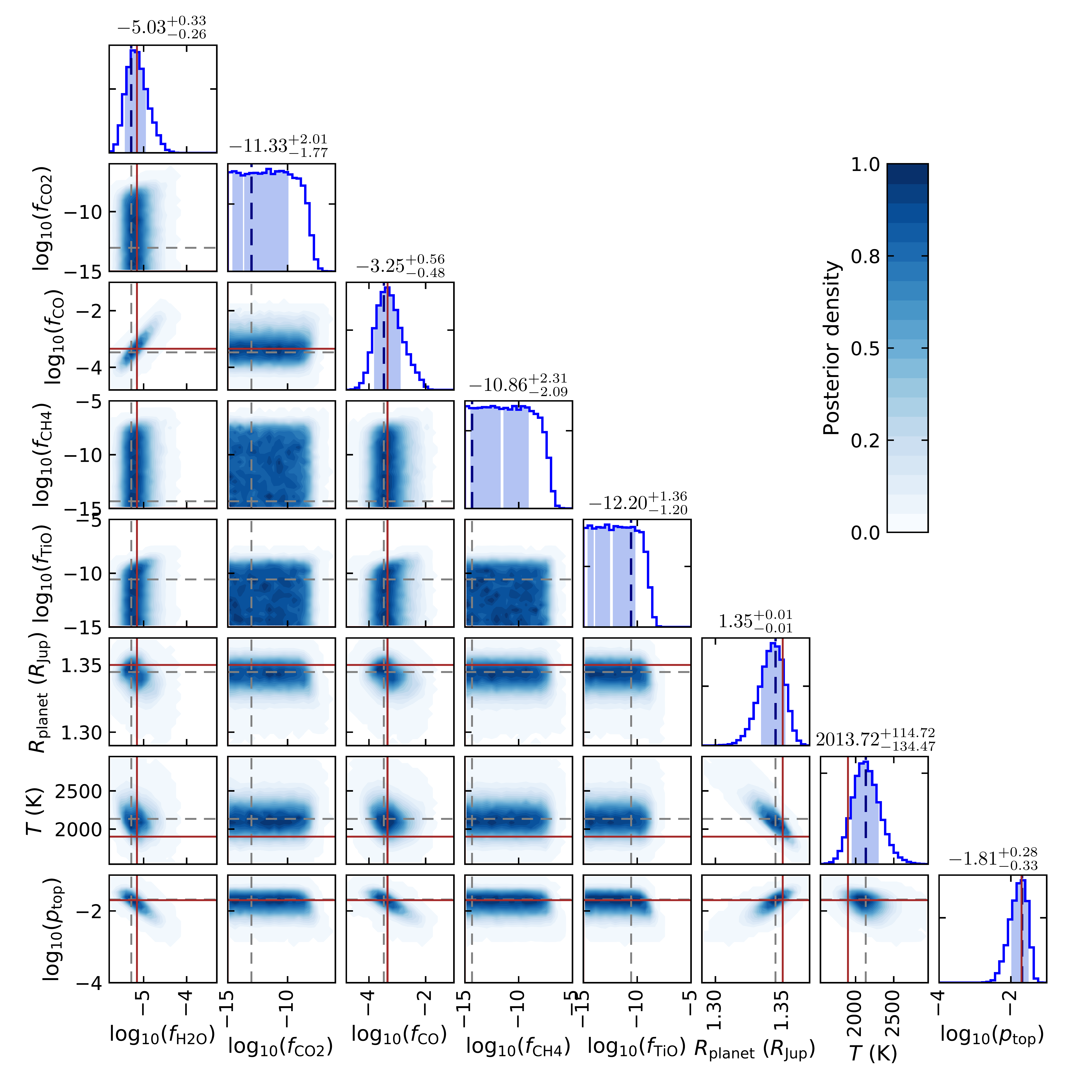}
\caption{As Figure \ref{posterior_nem} but for \pyratbay. In this case, dark red lines indicate the input values and dashed lines the median retrieved values. Shading in the histograms shows the +/-$\sigma$ envelope, which is obtained from the 68\% highest-posterior-density credible region \citep[see Appendix in][]{CubillosEtal2017apjRednoise}.
}
\label{posterior_pyrat}
\end{figure*}

\begin{figure*}[!h]
\centering
\includegraphics[width=1.0\textwidth]{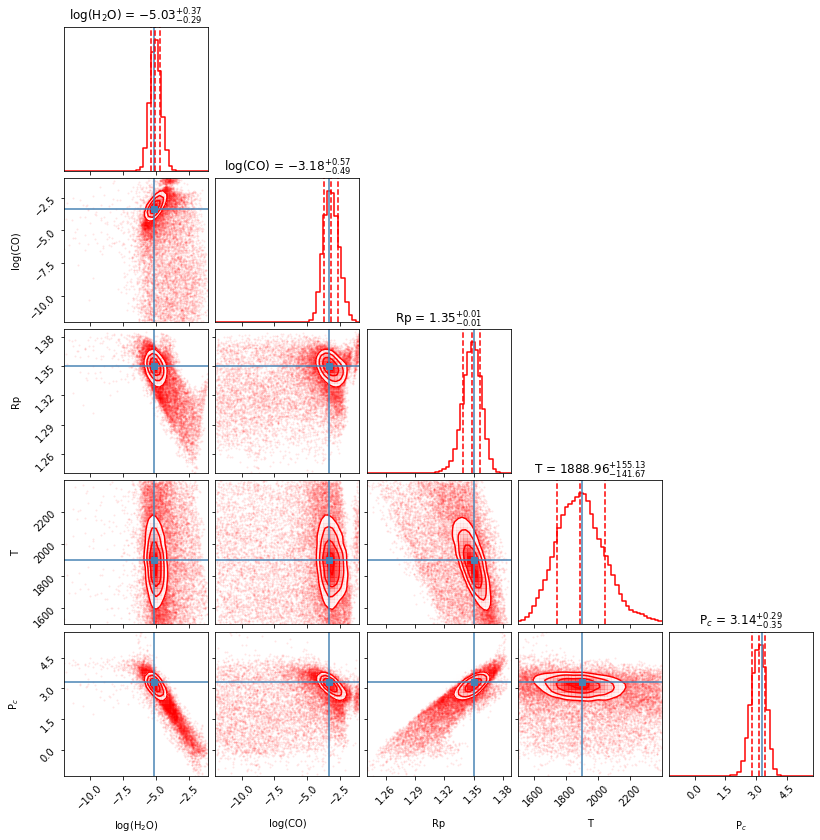}
\caption{As Figure \ref{posterior_nem} but for TauREx. Gases that were not included in the original model were not retrieved for in this case.}
\label{posterior_taurex}
\end{figure*}

\begin{figure*}[!h]
\centering
\includegraphics[width=1.0\textwidth]{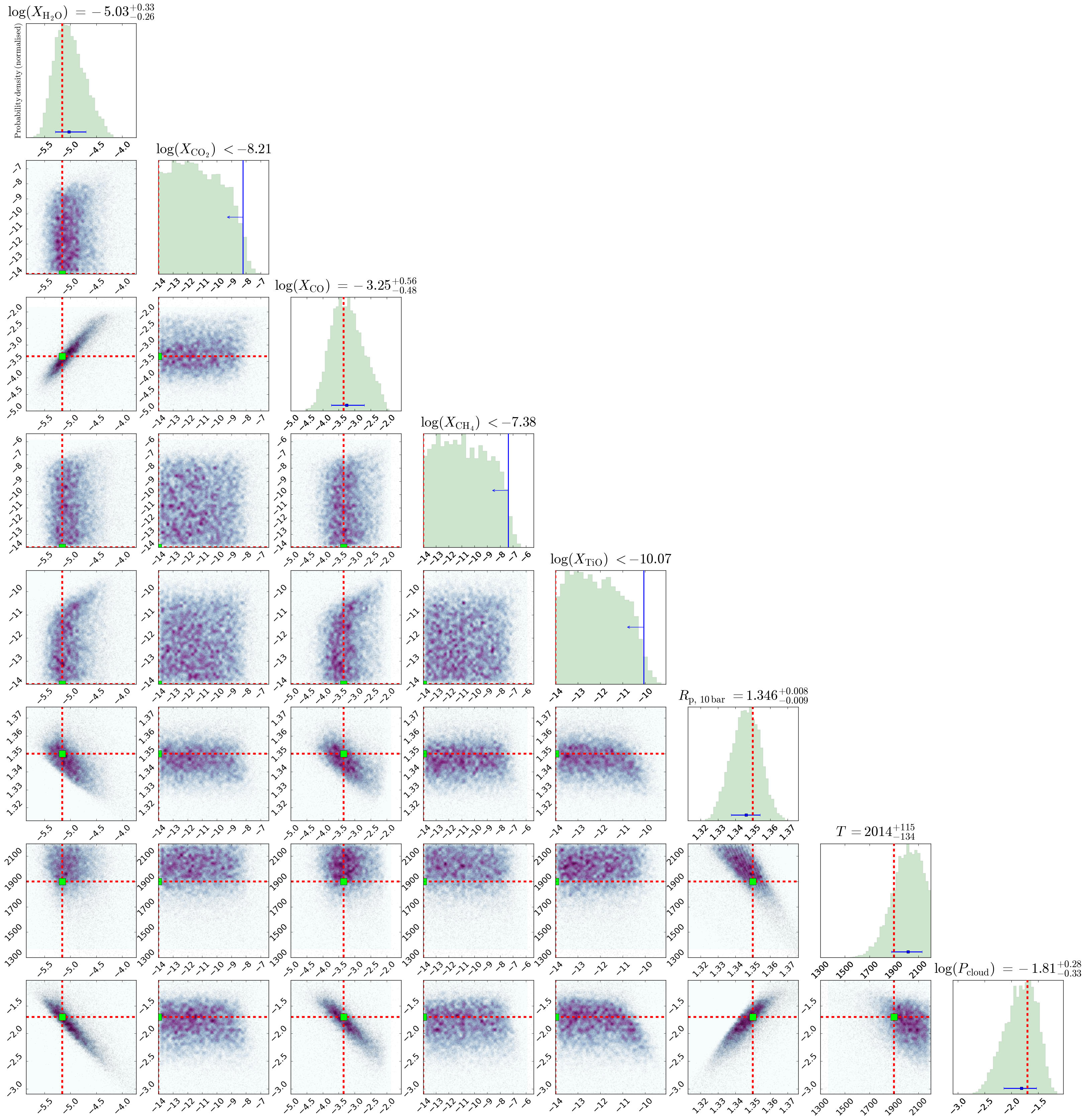}
\caption{As Figure \ref{posterior_nem} but for POSEIDON. Dashed red lines indicate input values. Blue points with error bars on the histograms show the median and $\pm 1\,\sigma$ confidence region. Where only an upper limit is retrieved, a blue line with an arrow indicates the $2\,\sigma$ limit.}
\label{posterior_poseidon}
\end{figure*}

To illustrate this, we show the full retrieved posteriors from each code for Planet 2 (Figures~\ref{posterior_nem}---~\ref{posterior_poseidon}). This planet was chosen as it is cloudy, allowing the effect of clouds on retrievals (and especially on parameter correlations) to be seen. Corner plots for NEMESIS, TauREx and ARCiS were generated using the corner.py routine \citep{corner}. 

All retrieved posteriors show that the abundances of the constrained gases H$_2$O and CO are inversely correlated with the cloud top pressure. Lower cloud top pressures correspond to cloud that sits higher in the atmosphere. The higher the cloud is, the larger the fraction of the atmospheric features that are obscured, so more H$_2$O and CO are required to offset a higher cloud. 

By contrast, 10-bar radius and cloud top pressure are correlated, because a lower cloud top pressure (higher cloud) means that a smaller radius is required to fit the observed spectrum. 

H$_2$O and CO abundances are correlated. This is likely to be because the spectrum is dominated by absorption due to H$_2$O. If the H$_2$O abundance increases, more CO is required for the feature to stand out against the H$_2$O absorption.

Finally, temperature and radius are inversely correlated. This is because both affect the atmospheric scale height in a similar way. Scale height is proportional to temperature, and inversely proportional to the gravitational acceleration. The gravitational acceleration $g \propto r^{-2}$, so the scale height is proportional to the square of the radius. The variation in transit depth is even proportional to the radius times the scaleheight, so $\propto r^{3}$. An increase in radius can therefore be offset by a decrease in temperature, and vice versa.

\subsection{Effects of cross section grids}
\label{pyratbay_crosssection}
\begin{figure*}[!h]
\centering
\includegraphics[width=1.0\textwidth]{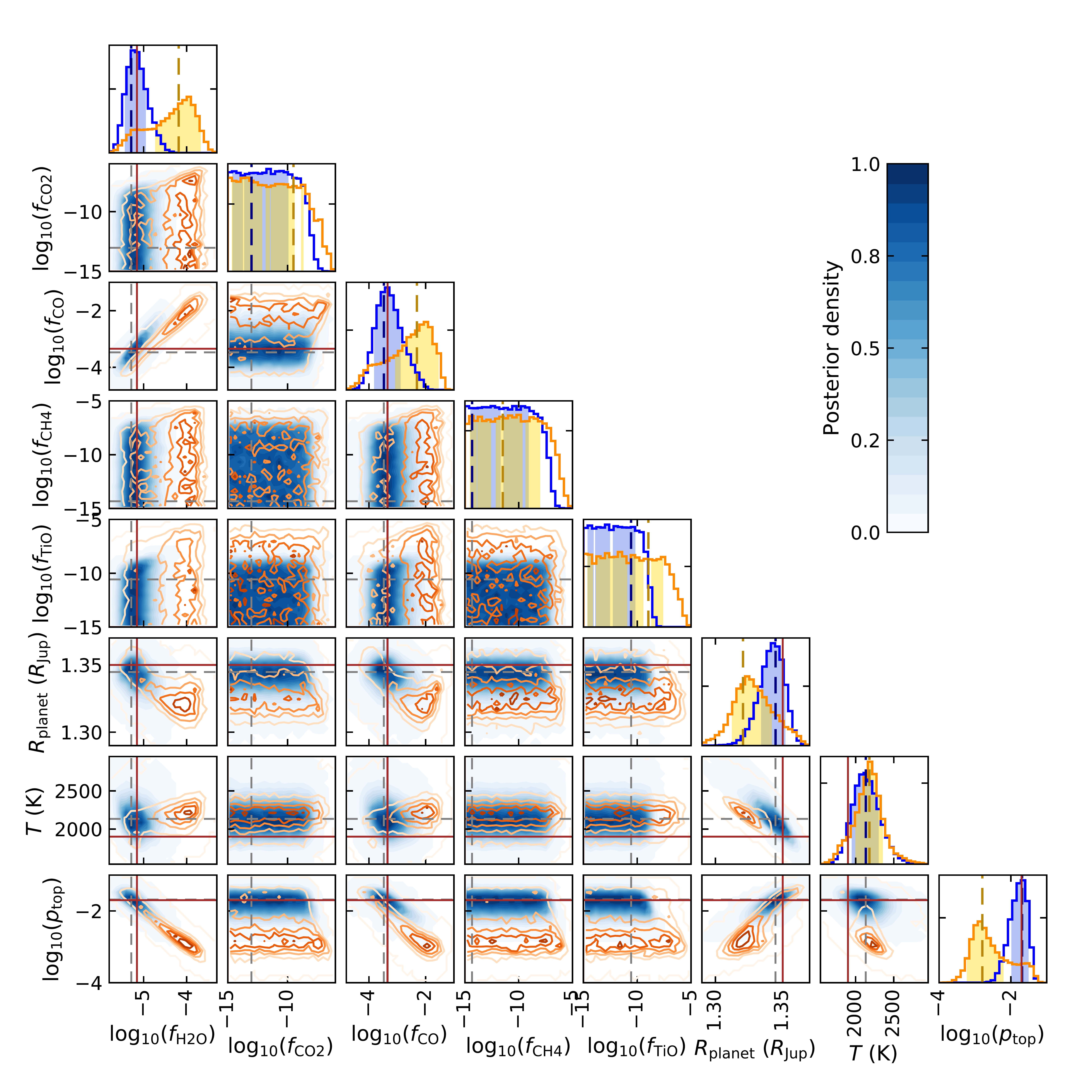}
\caption{As Figure \ref{posterior_nem} but for \pyratbay, with results plotted for both the original cross sections (yellow) and the TauREx cross sections (blue). Dark orange lines indicate the input values. Dashed lines indicate median and +/- $\sigma$ values. It is clear that the run using the same cross sections as used to generate the input spectrum provides a much better match to the true values. }
\label{posterior_cross_section}
\end{figure*}

The \pyratbay retrievals initially used different sampling for the gas absorption cross sections compared with the TauREx cross sections used the generate the input spectrum, which in the first iteration produced significantly discrepant results.  The main difference was that the {\pyratbay} cross sections sampled
the line transitions only up to 100 half-width at half maximum (HWHM)
away from the line center, whereas the TauREx cross sections sampled
the lines up to 500 HWHM or 25~cm$^{-1}$.  The discrepancies are
database-dependent (e.g., more significant for the CO molecule which
has more sparse line transitions) and are more significant at longer
wavelengths (due to the narrower Doppler line broadening). In Figure~\ref{posterior_cross_section} we show the effects of this on the retrieval. The {\pyratbay} run using the TauREx cross sections results in retrieved parameters that are much closer to the input values.
 Note that the difference in the cross sections arise only in the line sampling, as they both are computed from the same line lists.
 This is a good example of the way in which apparently minor variations in model set-up can affect retrieval outcomes.

\section{Conclusions}

We present a comparison of retrievals conducted by five different codes that provide overall very good agreement between them. We show that the parameters and uncertainties derived by these different codes are all comparable.

One important aspect to note is that small differences in the forward model setup can lead to noticable differences in the retrieval outcome. These systematic errors have to be considered when interpreting the absolute values of retrieval results, even though we show here that they are generally small.

\section{Outlook}

Ensuring model accuracy and completeness is key for developing tools that can be used to interpret data from missions such as Ariel. We have presented an example where line data tabulation substantially affected the accuracy of a retrieval; however, line data is just one aspect of modelling. 

So far, we have adopted very simple treatments of atmospheric thermal structure and clouds, and further work is needed to fully investigate these effects; some progress has already been made on cloud parameterisation for transit spectra (see e.g. \citealt{barstow20b} and \citealt{mai19}), and more sophisticated temperature parameterisations are already being applied to existing data. In addition, we have made the assumption here that the terminator of each planet is homogeneous, which we know is unlikely to be the case. Work by \cite{line16} and more recently by \cite{caldas2019, macdonald20, skaf2020, pluriel2020} demonstrate the ways in which this assumption can introduce bias into retrieval solutions. 

Comparative retrieval studies such as this one are required to understand the impacts of model differences on results, and new investigations with increased model complexity will certainly be required. Similar efforts for secondary eclipse spectra are also necessary and will form the subject of future studies by the Ariel Spectral Retrievals Working Group.

\begin{acknowledgements}
We thank Patrick Irwin for the use of NEMESIS. JKB was supported by a Royal Astronomical Society Research Fellowship during this project. This project has received funding from the European Research Council (ERC) under the European Union's Horizon 2020 research and innovation programme (grant agreement No 758892, ExoAI and grant agreement No 776403, ExoplANETS-A) and under the European Union's Seventh Framework Programme (FP7/2007-2013)/ ERC grant agreement numbers 617119 (ExoLights). Furthermore, we acknowledge funding by the Science and Technology Funding Council (STFC) grants: ST/K502406/1, ST/P000282/1, ST/P002153/1 and ST/S002634/1.
We acknowledge the availability and support from the High Performance Computing platforms (HPC) DIRAC and OzSTAR, which provided the computing resources necessary to perform this work.
\end{acknowledgements}

%
%

\bibliographystyle{spbasic}      
\bibliography{bibliography.bib}

\begin{thebibliography}{85}
\providecommand{\natexlab}[1]{#1}
\providecommand{\url}[1]{{#1}}
\providecommand{\urlprefix}{URL }
\expandafter\ifx\csname urlstyle\endcsname\relax
  \providecommand{\doi}[1]{DOI~\discretionary{}{}{}#1}\else
  \providecommand{\doi}{DOI~\discretionary{}{}{}\begingroup
  \urlstyle{rm}\Url}\fi
\providecommand{\eprint}[2][]{\url{#2}}

\bibitem[{{Ackerman} and {Marley}(2001)}]{ackerman01}
{Ackerman} AS, {Marley} MS (2001) {Precipitating Condensation Clouds in
  Substellar Atmospheres}. \apj 556:872--884, \doi{10.1086/321540},
  \eprint{astro-ph/0103423}

\bibitem[{{Al-Refaie} et~al.(2019){Al-Refaie}, {Changeat}, {Waldmann}, and
  {Tinetti}}]{alrefaie19}
{Al-Refaie} AF, {Changeat} Q, {Waldmann} IP, {Tinetti} G (2019) {TauREx III: A
  fast, dynamic and extendable framework for retrievals}. arXiv e-prints
  arXiv:1912.07759, \eprint{1912.07759}

\bibitem[{Arcangeli et~al.(2019)Arcangeli, Désert, Parmentier, Stevenson,
  Bean, Line, Kreidberg, Fortney, and Showman}]{arcangeli2019}
Arcangeli J, Désert JM, Parmentier V, Stevenson KB, Bean JL, Line MR,
  Kreidberg L, Fortney JJ, Showman AP (2019) Climate of an ultra hot jupiter.
  Astronomy \& Astrophysics 625:A136, \doi{10.1051/0004-6361/201834891},
  \urlprefix\url{http://dx.doi.org/10.1051/0004-6361/201834891}

\bibitem[{{Barstow}(2020)}]{barstow20b}
{Barstow} JK (2020) {Unveiling cloudy exoplanets: the influence of cloud model
  choices on retrieval solutions}. arXiv e-prints arXiv:2002.02945,
  \eprint{2002.02945}

\bibitem[{{Barstow} et~al.(2014){Barstow}, {Aigrain}, {Irwin}, {Hackler},
  {Fletcher}, {Lee}, and {Gibson}}]{barstow14}
{Barstow} JK, {Aigrain} S, {Irwin} PGJ, {Hackler} T, {Fletcher} LN, {Lee} JM,
  {Gibson} NP (2014) {Clouds on the Hot Jupiter HD189733b: Constraints from the
  Reflection Spectrum}. \apj 786:154, \doi{10.1088/0004-637X/786/2/154},
  \eprint{1403.6664}

\bibitem[{{Barstow} et~al.(2017){Barstow}, {Aigrain}, {Irwin}, and
  {Sing}}]{barstow17}
{Barstow} JK, {Aigrain} S, {Irwin} PGJ, {Sing} DK (2017) {A Consistent
  Retrieval Analysis of 10 Hot Jupiters Observed in Transmission}. \apj 834:50,
  \doi{10.3847/1538-4357/834/1/50}, \eprint{1610.01841}

\bibitem[{{Barstow} et~al.(2020){Barstow}, {Changeat}, {Garland}, {Line},
  {Rocchetto}, and {Waldmann}}]{barstow20}
{Barstow} JK, {Changeat} Q, {Garland} R, {Line} MR, {Rocchetto} M, {Waldmann}
  IP (2020) {A comparison of exoplanet spectroscopic retrieval tools}. arXiv
  e-prints arXiv:2002.01063, \eprint{2002.01063}

\bibitem[{Benneke(2015)}]{benneke2015}
Benneke B (2015) Strict upper limits on the carbon-to-oxygen ratios of eight
  hot jupiters from self-consistent atmospheric retrieval. \eprint{1504.07655}

\bibitem[{{Benneke} et~al.(2019){Benneke}, {Wong}, {Piaulet}, {Knutson},
  {Lothringer}, {Morley}, {Crossfield}, {Gao}, {Greene}, {Dressing},
  {Dragomir}, {Howard}, {McCullough}, {Kempton}, {Fortney}, and
  {Fraine}}]{benneke19}
{Benneke} B, {Wong} I, {Piaulet} C, {Knutson} HA, {Lothringer} J, {Morley} CV,
  {Crossfield} IJM, {Gao} P, {Greene} TP, {Dressing} C, {Dragomir} D, {Howard}
  AW, {McCullough} PR, {Kempton} EMR, {Fortney} JJ, {Fraine} J (2019) {Water
  Vapor and Clouds on the Habitable-zone Sub-Neptune Exoplanet K2-18b}. \apjl
  887(1):L14, \doi{10.3847/2041-8213/ab59dc}, \eprint{1909.04642}

\bibitem[{{Borysow}(2002{\natexlab{a}})}]{Borysow2002jqsrtH2H2lowT}
{Borysow} A (2002{\natexlab{a}}) {Collision-induced absorption coefficients of
  H$_{2}$ pairs at temperatures from 60 K to 1000 K}. \aap 390:779--782,
  \doi{10.1051/0004-6361:20020555}

\bibitem[{{Borysow}(2002{\natexlab{b}})}]{borysow02}
{Borysow} A (2002{\natexlab{b}}) {Collision-induced absorption coefficients of
  H$_{2}$ pairs at temperatures from 60 K to 1000 K}. Astronomy and
  Astrophysics 390:779--782, \doi{10.1051/0004-6361:20020555}

\bibitem[{{Borysow} and
  {Frommhold}(1989)}]{BorysowFrommhold1989apjH2HeOvertones}
{Borysow} A, {Frommhold} L (1989) {Collision-induced Infrared Spectra of H 2-He
  Pairs at Temperatures from 18 to 7000 K. II. Overtone and Hot Bands}. \apj
  341:549, \doi{10.1086/167515}

\bibitem[{{Borysow} et~al.(1989){Borysow}, {Frommhold}, and
  {Moraldi}}]{BorysowEtal1989apjH2HeRVRT}
{Borysow} A, {Frommhold} L, {Moraldi} M (1989) {Collision-induced Infrared
  Spectra of H 2-He Pairs Involving 0 1 Vibrational Transitions and
  Temperatures from 18 to 7000 K}. \apj 336:495, \doi{10.1086/167027}

\bibitem[{{Borysow} et~al.(2001{\natexlab{a}}){Borysow}, {Jorgensen}, and
  {Fu}}]{BorysowEtal2001jqsrtH2H2highT}
{Borysow} A, {Jorgensen} UG, {Fu} Y (2001{\natexlab{a}}) {High-temperature
  (1000-7000 K) collision-induced absorption of H''2 pairs computed from the
  first principles, with application to cool and dense stellar atmospheres}.
  Journal of Quantitative Spectroscopy and Radiative Transfer 68:235--255,
  \doi{10.1016/S0022-4073(00)00023-6}

\bibitem[{{Borysow} et~al.(2001{\natexlab{b}}){Borysow}, {Jorgensen}, and
  {Fu}}]{Borysow2001}
{Borysow} A, {Jorgensen} UG, {Fu} Y (2001{\natexlab{b}}) {High-temperature
  (1000-7000 K) collision-induced absorption of H``2 pairs computed from the
  first principles, with application to cool and dense stellar atmospheres}.
  \jqsrt 68:235--255, \doi{10.1016/S0022-4073(00)00023-6}

\bibitem[{{Buchner} et~al.(2014){Buchner}, {Georgakakis}, {Nandra}, {Hsu},
  {Rangel}, {Brightman}, {Merloni}, {Salvato}, {Donley}, and
  {Kocevski}}]{buchner14}
{Buchner} J, {Georgakakis} A, {Nandra} K, {Hsu} L, {Rangel} C, {Brightman} M,
  {Merloni} A, {Salvato} M, {Donley} J, {Kocevski} D (2014) {X-ray spectral
  modelling of the AGN obscuring region in the CDFS: Bayesian model selection
  and catalogue}. \aap 564:A125, \doi{10.1051/0004-6361/201322971},
  \eprint{1402.0004}

\bibitem[{{Burrows} et~al.(2000){Burrows}, {Marley}, and
  {Sharp}}]{BurrowsEtal2000apjBDspectra}
{Burrows} A, {Marley} MS, {Sharp} CM (2000) {The Near-Infrared and Optical
  Spectra of Methane Dwarfs and Brown Dwarfs}. \apj 531:438--446,
  \doi{10.1086/308462}, \eprint{astro-ph/9908078}

\bibitem[{Caldas et~al.(2019)Caldas, Leconte, Selsis, Waldmann, Bordé,
  Rocchetto, and Charnay}]{caldas2019}
Caldas A, Leconte J, Selsis F, Waldmann IP, Bordé P, Rocchetto M, Charnay B
  (2019) Effects of a fully 3d atmospheric structure on exoplanet transmission
  spectra: retrieval biases due to day–night temperature gradients. Astronomy
  \& Astrophysics 623:A161, \doi{10.1051/0004-6361/201834384},
  \urlprefix\url{http://dx.doi.org/10.1051/0004-6361/201834384}

\bibitem[{{Changeat} et~al.(2019){Changeat}, {Edwards}, {Waldmann}, and
  {Tinetti}}]{changeat19}
{Changeat} Q, {Edwards} B, {Waldmann} IP, {Tinetti} G (2019) {Toward a More
  Complex Description of Chemical Profiles in Exoplanet Retrievals: A Two-layer
  Parameterization}. \apj 886(1):39, \doi{10.3847/1538-4357/ab4a14},
  \eprint{1903.11180}

\bibitem[{{Changeat} et~al.(2020){Changeat}, {Al-Refaie}, {Mugnai}, {Edwards},
  {Waldmann}, {Pascale}, and {Tinetti}}]{2020AJ....160...80C}
{Changeat} Q, {Al-Refaie} A, {Mugnai} LV, {Edwards} B, {Waldmann} IP, {Pascale}
  E, {Tinetti} G (2020) {Alfnoor: A Retrieval Simulation of the Ariel Target
  List}. \aj 160(2):80, \doi{10.3847/1538-3881/ab9a53}, \eprint{2003.01839}

\bibitem[{Chubb et~al.(2020)Chubb, Rocchetto, Al-Refaie, Waldmann, Min,
  Barstow, Molli{\'e}re, Phillips, Tennyson, and Yurchenko}]{chubb20}
Chubb KL, Rocchetto M, Al-Refaie AF, Waldmann I, Min M, Barstow J, Molli{\'e}re
  P, Phillips MW, Tennyson J, Yurchenko SN (2020) The {ExoMolOP Database:
  C}ross-sections and k-tables for molecules of interest in high-temperature
  exoplanet atmospheres. A\&A (In revision)

\bibitem[{Cubillos(2018)}]{cubillos2018}
Cubillos (2018) Pyratbay retrieval code.
  \urlprefix\url{https://pcubillos.github.io/pyratbay/index.html}

\bibitem[{{Cubillos} et~al.(2017){Cubillos}, {Harrington}, {Loredo}, {Lust},
  {Blecic}, and {Stemm}}]{CubillosEtal2017apjRednoise}
{Cubillos} P, {Harrington} J, {Loredo} TJ, {Lust} NB, {Blecic} J, {Stemm} M
  (2017) {On Correlated-noise Analyses Applied to Exoplanet Light Curves}. \aj
  153:3, \doi{10.3847/1538-3881/153/1/3}, \eprint{1610.01336}

\bibitem[{{Cubillos}(2017)}]{Cubillos2017apjCompress}
{Cubillos} PE (2017) {An Algorithm to Compress Line-transition Data for
  Radiative-transfer Calculations}. \apj 850:32,
  \doi{10.3847/1538-4357/aa9228}, \eprint{1710.02556}

\bibitem[{{Cubillos} and {Blecic}(2021)}]{CubillosBlecic2021mnrasPyratBay}
{Cubillos} PE, {Blecic} J (2021) {The {Pyrat Bay} Framework for Exoplanet
  Atmospheric Modeling: A Population Study of Hubble/WFC3 Transmission
  Spectra}. arXiv e-prints arXiv:2105.05598, \eprint{2105.05598}

\bibitem[{Demory et~al.(2016)Demory, Gillon, de~Wit, Madhusudhan, Bolmont,
  Heng, Kataria, Lewis, Hu, Krick, and et~al.}]{demory2016}
Demory BO, Gillon M, de~Wit J, Madhusudhan N, Bolmont E, Heng K, Kataria T,
  Lewis N, Hu R, Krick J, et~al (2016) A map of the large day–night
  temperature gradient of a super-earth exoplanet. Nature 532(7598):207–209,
  \doi{10.1038/nature17169},
  \urlprefix\url{http://dx.doi.org/10.1038/nature17169}

\bibitem[{{Feroz} et~al.(2009){Feroz}, {Hobson}, and {Bridges}}]{feroz09}
{Feroz} F, {Hobson} MP, {Bridges} M (2009) {MULTINEST: an efficient and robust
  Bayesian inference tool for cosmology and particle physics}. \mnras
  398:1601--1614, \doi{10.1111/j.1365-2966.2009.14548.x}, \eprint{0809.3437}

\bibitem[{{Fletcher} et~al.(2009){Fletcher}, {Orton}, {Teanby}, and
  {Irwin}}]{fletcher09}
{Fletcher} LN, {Orton} GS, {Teanby} NA, {Irwin} PGJ (2009) {Phosphine on
  Jupiter and Saturn from Cassini/CIRS}. \icarus 202:543--564,
  \doi{10.1016/j.icarus.2009.03.023}

\bibitem[{Foreman-Mackey(2016)}]{corner}
Foreman-Mackey D (2016) corner.py: Scatterplot matrices in python. The Journal
  of Open Source Software 1(2):24, \doi{10.21105/joss.00024},
  \urlprefix\url{https://doi.org/10.21105/joss.00024}

\bibitem[{{Gandhi} and {Madhusudhan}(2018)}]{gandhi2018}
{Gandhi} S, {Madhusudhan} N (2018) {Retrieval of exoplanet emission spectra
  with HyDRA}. \mnras 474(1):271--288, \doi{10.1093/mnras/stx2748},
  \eprint{1710.06433}

\bibitem[{{Gelman} and {Rubin}(1992)}]{GelmanRubin1992}
{Gelman} A, {Rubin} DB (1992) Inference from iterative simulation using
  multiple sequences. Statistical Science 7:457--511

\bibitem[{{Goody} and {Yung}(1989)}]{goodyyung}
{Goody} RM, {Yung} YL (1989) {Atmospheric radiation : theoretical basis}

\bibitem[{{Gordon} et~al.(2016){Gordon}, {Rothman}, {Wilzewski}, {Kochanov},
  {Hill}, {Tan}, and {Wcislo}}]{gordon16}
{Gordon} I, {Rothman} LS, {Wilzewski} JS, {Kochanov} RV, {Hill} C, {Tan} Y,
  {Wcislo} P (2016) {HITRAN2016 : new and improved data and tools towards
  studies of planetary atmospheres}. In: AAS/Division for Planetary Sciences
  Meeting Abstracts \#48, AAS/Division for Planetary Sciences Meeting
  Abstracts, p 421.13

\bibitem[{{Gordon} et~al.(2017){Gordon}, {Rothman}, {Hill}, {Kochanov}, {Tan},
  {Bernath}, {Birk}, {Boudon}, {Campargue}, {Chance}, {Drouin}, {Flaud},
  {Gamache}, {Hodges}, {Jacquemart}, {Perevalov}, {Perrin}, {Shine}, {Smith},
  {Tennyson}, {Toon}, {Tran}, {Tyuterev}, {Barbe}, {Cs{\'a}sz{\'a}r}, {Devi},
  {Furtenbacher}, {Harrison}, {Hartmann}, {Jolly}, {Johnson}, {Karman},
  {Kleiner}, {Kyuberis}, {Loos}, {Lyulin}, {Massie}, {Mikhailenko},
  {Moazzen-Ahmadi}, {M{\"u}ller}, {Naumenko}, {Nikitin}, {Polyansky}, {Rey},
  {Rotger}, {Sharpe}, {Sung}, {Starikova}, {Tashkun}, {Auwera}, {Wagner},
  {Wilzewski}, {Wcis{\l}o}, {Yu}, and {Zak}}]{GordonEtal2017jqsrtHITRAN2016}
{Gordon} IE, {Rothman} LS, {Hill} C, {Kochanov} RV, {Tan} Y, {Bernath} PF,
  {Birk} M, {Boudon} V, {Campargue} A, {Chance} KV, {Drouin} BJ, {Flaud} JM,
  {Gamache} RR, {Hodges} JT, {Jacquemart} D, {Perevalov} VI, {Perrin} A,
  {Shine} KP, {Smith} MAH, {Tennyson} J, {Toon} GC, {Tran} H, {Tyuterev} VG,
  {Barbe} A, {Cs{\'a}sz{\'a}r} AG, {Devi} VM, {Furtenbacher} T, {Harrison} JJ,
  {Hartmann} JM, {Jolly} A, {Johnson} TJ, {Karman} T, {Kleiner} I, {Kyuberis}
  AA, {Loos} J, {Lyulin} OM, {Massie} ST, {Mikhailenko} SN, {Moazzen-Ahmadi} N,
  {M{\"u}ller} HSP, {Naumenko} OV, {Nikitin} AV, {Polyansky} OL, {Rey} M,
  {Rotger} M, {Sharpe} SW, {Sung} K, {Starikova} E, {Tashkun} SA, {Auwera} JVe,
  {Wagner} G, {Wilzewski} J, {Wcis{\l}o} P, {Yu} S, {Zak} EJ (2017) {The
  HITRAN2016 molecular spectroscopic database}. \jqsrt 203:3--69,
  \doi{10.1016/j.jqsrt.2017.06.038}

\bibitem[{{Harrington}(2016)}]{harrington16}
{Harrington} J (2016) {Atmospheric Retrievals from Exoplanet Observations and
  Simulations with BART}. NASA Proposal id.16-XPR16-10

\bibitem[{{Hoeijmakers} et~al.(2018){Hoeijmakers}, {Ehrenreich}, {Heng},
  {Kitzmann}, {Grimm}, {Allart}, {Deitrick}, {Wyttenbach}, {Oreshenko}, and
  {Pino}}]{hoeijmakers18}
{Hoeijmakers} HJ, {Ehrenreich} D, {Heng} K, {Kitzmann} D, {Grimm} SL, {Allart}
  R, {Deitrick} R, {Wyttenbach} A, {Oreshenko} M, {Pino} L (2018) {Atomic iron
  and titanium in the atmosphere of the exoplanet KELT-9b}. \nat
  560(7719):453--455, \doi{10.1038/s41586-018-0401-y}, \eprint{1808.05653}

\bibitem[{Huang et~al.(2017)Huang, Schwenke, Freedman, and Lee}]{huang17}
Huang X, Schwenke DW, Freedman RS, Lee TJ (2017) Ames-2016 line lists for 13
  isotopologues of co2: Updates, consistency, and remaining issues. Journal of
  Quantitative Spectroscopy and Radiative Transfer 203:224 -- 241,
  \doi{https://doi.org/10.1016/j.jqsrt.2017.04.026},
  \urlprefix\url{http://www.sciencedirect.com/science/article/pii/S0022407317300547},
  hITRAN2016 Special Issue

\bibitem[{{Irwin} et~al.(2008){Irwin}, {Teanby}, {de Kok}, {Fletcher},
  {Howett}, {Tsang}, {Wilson}, {Calcutt}, {Nixon}, and {Parrish}}]{irwin08}
{Irwin} PGJ, {Teanby} NA, {de Kok} R, {Fletcher} LN, {Howett} CJA, {Tsang} CCC,
  {Wilson} CF, {Calcutt} SB, {Nixon} CA, {Parrish} PD (2008) {The NEMESIS
  planetary atmosphere radiative transfer and retrieval tool}. JQSRT
  109:1136--1150, \doi{10.1016/j.jqsrt.2007.11.006}

\bibitem[{{Karman} et~al.(2019){Karman}, {Gordon}, {van der Avoird}, {Baranov},
  {Boulet}, {Drouin}, {Groenenboom}, {Gustafsson}, {Hartmann}, {Kurucz},
  {Rothman}, {Sun}, {Sung}, {Thalman}, {Tran}, {Wishnow}, {Wordsworth},
  {Vigasin}, {Volkamer}, and {van der Zande}}]{karman2019}
{Karman} T, {Gordon} IE, {van der Avoird} A, {Baranov} YI, {Boulet} C, {Drouin}
  BJ, {Groenenboom} GC, {Gustafsson} M, {Hartmann} JM, {Kurucz} RL, {Rothman}
  LS, {Sun} K, {Sung} K, {Thalman} R, {Tran} H, {Wishnow} EH, {Wordsworth} R,
  {Vigasin} AA, {Volkamer} R, {van der Zande} WJ (2019) {Update of the HITRAN
  collision-induced absorption section}. \icarus 328:160--175,
  \doi{10.1016/j.icarus.2019.02.034}

\bibitem[{{Kitzmann} et~al.(2019){Kitzmann}, {Heng}, {Oreshenko}, {Grimm},
  {Apai}, {Bowler}, {Burgasser}, and {Marley}}]{kitzmann2019}
{Kitzmann} D, {Heng} K, {Oreshenko} M, {Grimm} SL, {Apai} D, {Bowler} BP,
  {Burgasser} AJ, {Marley} MS (2019) {Helios-r.2 -- A new Bayesian, open-source
  retrieval model for brown dwarfs and exoplanet atmospheres}. arXiv e-prints
  arXiv:1910.01070, \eprint{1910.01070}

\bibitem[{{Krissansen-Totton} et~al.(2018){Krissansen-Totton}, {Garland},
  {Irwin}, and {Catling}}]{krissansen-totton18}
{Krissansen-Totton} J, {Garland} R, {Irwin} P, {Catling} DC (2018)
  {Detectability of Biosignatures in Anoxic Atmospheres with the James Webb
  Space Telescope: A TRAPPIST-1e Case Study}. \aj 156:114,
  \doi{10.3847/1538-3881/aad564}, \eprint{1808.08377}

\bibitem[{{Kurucz}(1970)}]{Kurucz1970saorsAtlas}
{Kurucz} RL (1970) {Atlas: a Computer Program for Calculating Model Stellar
  Atmospheres}. SAO Special Report 309

\bibitem[{{Lacis} and {Oinas}(1991)}]{lacis91}
{Lacis} AA, {Oinas} V (1991) {A description of the correlated-k distribution
  method for modelling nongray gaseous absorption, thermal emission, and
  multiple scattering in vertically inhomogeneous atmospheres}. J Geophys Res
  96:9027--9064, \doi{10.1029/90JD01945}

\bibitem[{{Lavie} et~al.(2017){Lavie}, {Mendon{\c c}a}, {Mordasini}, {Malik},
  {Bonnefoy}, {Demory}, {Oreshenko}, {Grimm}, {Ehrenreich}, and
  {Heng}}]{lavie17}
{Lavie} B, {Mendon{\c c}a} JM, {Mordasini} C, {Malik} M, {Bonnefoy} M, {Demory}
  BO, {Oreshenko} M, {Grimm} SL, {Ehrenreich} D, {Heng} K (2017)
  {HELIOS-RETRIEVAL: An Open-source, Nested Sampling Atmospheric Retrieval
  Code; Application to the HR 8799 Exoplanets and Inferred Constraints for
  Planet Formation}. \aj 154:91, \doi{10.3847/1538-3881/aa7ed8},
  \eprint{1610.03216}

\bibitem[{{Lecavelier Des Etangs} et~al.(2008){Lecavelier Des Etangs}, {Pont},
  {Vidal-Madjar}, and {Sing}}]{LecavelierDesEtangsEtal2008aaRayleighHD189}
{Lecavelier Des Etangs} A, {Pont} F, {Vidal-Madjar} A, {Sing} D (2008)
  {Rayleigh scattering in the transit spectrum of HD 189733b}. \aap
  481:L83--L86, \doi{10.1051/0004-6361:200809388}, \eprint{0802.3228}

\bibitem[{{Lee} et~al.(2012){Lee}, {Fletcher}, and {Irwin}}]{lee12}
{Lee} JM, {Fletcher} LN, {Irwin} PGJ (2012) {Optimal estimation retrievals of
  the atmospheric structure and composition of HD 189733b from secondary
  eclipse spectroscopy}. \mnras 420:170--182,
  \doi{10.1111/j.1365-2966.2011.20013.x}, \eprint{1110.2934}

\bibitem[{{Li} et~al.(2015){Li}, {Gordon}, {Rothman}, {Tan}, {Hu}, {Kassi},
  {Campargue}, and {Medvedev}}]{li15}
{Li} G, {Gordon} IE, {Rothman} LS, {Tan} Y, {Hu} SM, {Kassi} S, {Campargue} A,
  {Medvedev} ES (2015) {Rovibrational Line Lists for Nine Isotopologues of the
  CO Molecule in the X $^{1}${\ensuremath{\Sigma}}$^{+}$ Ground Electronic
  State}. The Astrophysical Journal Supplement Series 216:15,
  \doi{10.1088/0067-0049/216/1/15}

\bibitem[{{Line} and {Parmentier}(2016)}]{line16}
{Line} MR, {Parmentier} V (2016) {The Influence of Nonuniform Cloud Cover on
  Transit Transmission Spectra}. \apj 820:78, \doi{10.3847/0004-637X/820/1/78},
  \eprint{1511.09443}

\bibitem[{{Line} et~al.(2013){Line}, {Wolf}, {Zhang}, {Knutson}, {Kammer},
  {Ellison}, {Deroo}, {Crisp}, and {Yung}}]{line13a}
{Line} MR, {Wolf} AS, {Zhang} X, {Knutson} H, {Kammer} JA, {Ellison} E, {Deroo}
  P, {Crisp} D, {Yung} YL (2013) {A Systematic Retrieval Analysis of Secondary
  Eclipse Spectra. I. A Comparison of Atmospheric Retrieval Techniques}. ApJ
  775:137, \doi{10.1088/0004-637X/775/2/137}, \eprint{1304.5561}

\bibitem[{{MacDonald} and {Madhusudhan}(2017)}]{macdonald17}
{MacDonald} RJ, {Madhusudhan} N (2017) {HD 209458b in new light: evidence of
  nitrogen chemistry, patchy clouds and sub-solar water}. \mnras
  469(2):1979--1996, \doi{10.1093/mnras/stx804}, \eprint{1701.01113}

\bibitem[{{MacDonald} et~al.(2020){MacDonald}, {Goyal}, and
  {Lewis}}]{macdonald20}
{MacDonald} RJ, {Goyal} JM, {Lewis} NK (2020) {Why Is it So Cold in Here?
  Explaining the Cold Temperatures Retrieved from Transmission Spectra of
  Exoplanet Atmospheres}. \apjl 893(2):L43, \doi{10.3847/2041-8213/ab8238},
  \eprint{2003.11548}

\bibitem[{{Madhusudhan} and {Seager}(2009)}]{madhusudhan2009}
{Madhusudhan} N, {Seager} S (2009) {A Temperature and Abundance Retrieval
  Method for Exoplanet Atmospheres}. ApJ 707:24--39,
  \doi{10.1088/0004-637X/707/1/24}, \eprint{0910.1347}

\bibitem[{{Mai} and {Line}(2019)}]{mai19}
{Mai} C, {Line} MR (2019) {Exploring Exoplanet Cloud Assumptions in JWST
  Transmission Spectra}. \apj 883(2):144, \doi{10.3847/1538-4357/ab3e6d},
  \eprint{1908.10904}

\bibitem[{McKemmish et~al.(2019)McKemmish, Masseron, Hoeijmakers, Pérez-Mesa,
  Grimm, Yurchenko, and Tennyson}]{mckemmish19}
McKemmish LK, Masseron T, Hoeijmakers HJ, Pérez-Mesa V, Grimm SL, Yurchenko
  SN, Tennyson J (2019) {ExoMol molecular line lists – XXXIII. The spectrum
  of Titanium Oxide}. Monthly Notices of the Royal Astronomical Society
  488(2):2836--2854

\bibitem[{{Min} et~al.(2005){Min}, {Hovenier}, and {de
  Koter}}]{2005A&A...432..909M}
{Min} M, {Hovenier} JW, {de Koter} A (2005) {Modeling optical properties of
  cosmic dust grains using a distribution of hollow spheres}. \aap
  432(3):909--920, \doi{10.1051/0004-6361:20041920}, \eprint{astro-ph/0503068}

\bibitem[{{Min} et~al.(2020){Min}, {Ormel}, {Chubb}, {Helling}, and
  {Kawashima}}]{min20}
{Min} M, {Ormel} CW, {Chubb} K, {Helling} C, {Kawashima} Y (2020) {The ARCiS
  framework for Exoplanet Atmospheres: Modelling Philosophy and Retrieval}.
  arXiv e-prints arXiv:2006.12821, \eprint{2006.12821}

\bibitem[{{Molli{\`e}re} et~al.(2020){Molli{\`e}re}, {Stolker}, {Lacour},
  {Otten}, {Shangguan}, {Charnay}, {Molyarova}, {Nowak}, {Henning}, {Marleau},
  {Semenov}, {van Dishoeck}, {Eisenhauer}, {Garcia}, {Garcia Lopez}, {Girard},
  {Greenbaum}, {Hinkley}, {Kervella}, {Kreidberg}, {Maire}, {Nasedkin},
  {Pueyo}, {Snellen}, {Vigan}, {Wang}, {de Zeeuw}, and
  {Zurlo}}]{2020A&A...640A.131M}
{Molli{\`e}re} P, {Stolker} T, {Lacour} S, {Otten} GPPL, {Shangguan} J,
  {Charnay} B, {Molyarova} T, {Nowak} M, {Henning} T, {Marleau} GD, {Semenov}
  DA, {van Dishoeck} E, {Eisenhauer} F, {Garcia} P, {Garcia Lopez} R, {Girard}
  JH, {Greenbaum} AZ, {Hinkley} S, {Kervella} P, {Kreidberg} L, {Maire} AL,
  {Nasedkin} E, {Pueyo} L, {Snellen} IAG, {Vigan} A, {Wang} J, {de Zeeuw} PT,
  {Zurlo} A (2020) {Retrieving scattering clouds and disequilibrium chemistry
  in the atmosphere of HR 8799e}. \aap 640:A131,
  \doi{10.1051/0004-6361/202038325}, \eprint{2006.09394}

\bibitem[{Mollière et~al.(2019)Mollière, Wardenier, van Boekel, Henning,
  Molaverdikhani, and Snellen}]{Mollire19}
Mollière P, Wardenier JP, van Boekel R, Henning T, Molaverdikhani K, Snellen
  IAG (2019) petitradtrans. Astronomy \& Astrophysics 627:A67,
  \doi{10.1051/0004-6361/201935470},
  \urlprefix\url{http://dx.doi.org/10.1051/0004-6361/201935470}

\bibitem[{{Mugnai} et~al.(2020){Mugnai}, {Pascale}, {Edwards}, {Papageorgiou},
  and {Sarkar}}]{mugnai19}
{Mugnai} LV, {Pascale} E, {Edwards} B, {Papageorgiou} A, {Sarkar} S (2020)
  {ArielRad: the Ariel radiometric model}. Experimental Astronomy
  50(2-3):303--328, \doi{10.1007/s10686-020-09676-7}, \eprint{2009.07824}

\bibitem[{{Ormel} and {Min}(2019{\natexlab{a}})}]{ormel19}
{Ormel} CW, {Min} M (2019{\natexlab{a}}) {ARCiS framework for exoplanet
  atmospheres. The cloud transport model}. \aap 622:A121,
  \doi{10.1051/0004-6361/201833678}, \eprint{1812.05053}

\bibitem[{{Ormel} and {Min}(2019{\natexlab{b}})}]{2019A&A...622A.121O}
{Ormel} CW, {Min} M (2019{\natexlab{b}}) {ARCiS framework for exoplanet
  atmospheres. The cloud transport model}. \aap 622:A121,
  \doi{10.1051/0004-6361/201833678}, \eprint{1812.05053}

\bibitem[{{Pascale} et~al.(2018){Pascale}, {Bezawada}, {Barstow}, {Beaulieu},
  {Bowles}, {Coud{\'e} du Foresto}, {Coustenis}, {Decin}, {Drossart},
  {Eccleston}, {Encrenaz}, {Forget}, {Griffin}, {G{\"u}del}, {Hartogh},
  {Heske}, {Lagage}, {Leconte}, {Malaguti}, {Micela}, {Middleton}, {Min},
  {Moneti}, {Morales}, {Mugnai}, {Ollivier}, {Pace}, {Papageorgiou},
  {Pilbratt}, {Puig}, {Rataj}, {Ray}, {Ribas}, {Rocchetto}, {Sarkar}, {Selsis},
  {Taylor}, {Tennyson}, {Tinetti}, {Turrini}, {Vandenbussche}, {Venot},
  {Waldmann}, {Wolkenberg}, {Wright}, {Zapatero Osorio}, and
  {Zingales}}]{pascale18}
{Pascale} E, {Bezawada} N, {Barstow} J, {Beaulieu} JP, {Bowles} N, {Coud{\'e}
  du Foresto} V, {Coustenis} A, {Decin} L, {Drossart} P, {Eccleston} P,
  {Encrenaz} T, {Forget} F, {Griffin} M, {G{\"u}del} M, {Hartogh} P, {Heske} A,
  {Lagage} PO, {Leconte} J, {Malaguti} P, {Micela} G, {Middleton} K, {Min} M,
  {Moneti} A, {Morales} JC, {Mugnai} L, {Ollivier} M, {Pace} E, {Papageorgiou}
  A, {Pilbratt} G, {Puig} L, {Rataj} M, {Ray} T, {Ribas} I, {Rocchetto} M,
  {Sarkar} S, {Selsis} F, {Taylor} W, {Tennyson} J, {Tinetti} G, {Turrini} D,
  {Vandenbussche} B, {Venot} O, {Waldmann} IP, {Wolkenberg} P, {Wright} G,
  {Zapatero Osorio} MR, {Zingales} T (2018) {The ARIEL space mission}. In:
  \procspie, Society of Photo-Optical Instrumentation Engineers (SPIE)
  Conference Series, vol 10698, p 106980H, \doi{10.1117/12.2311838}

\bibitem[{{Pinhas} et~al.(2019){Pinhas}, {Madhusudhan}, {Gandhi}, and
  {MacDonald}}]{pinhas19}
{Pinhas} A, {Madhusudhan} N, {Gandhi} S, {MacDonald} R (2019) {H$_{2}$O
  abundances and cloud properties in ten hot giant exoplanets}. \mnras
  482:1485--1498, \doi{10.1093/mnras/sty2544}, \eprint{1811.00011}

\bibitem[{Pluriel et~al.(2020)Pluriel, Zingales, Leconte, and
  Parmentier}]{pluriel2020}
Pluriel W, Zingales T, Leconte J, Parmentier V (2020) Strong biases in
  retrieved atmospheric composition caused by day–night chemical
  heterogeneities. Astronomy \& Astrophysics 636:A66,
  \doi{10.1051/0004-6361/202037678},
  \urlprefix\url{http://dx.doi.org/10.1051/0004-6361/202037678}

\bibitem[{{Polyansky} et~al.(2018){Polyansky}, {Kyuberis}, {Zobov}, {Tennyson},
  {Yurchenko}, and {Lodi}}]{pokazatel}
{Polyansky} OL, {Kyuberis} AA, {Zobov} NF, {Tennyson} J, {Yurchenko} SN, {Lodi}
  L (2018) {ExoMol molecular line lists XXX: a complete high-accuracy line list
  for water}. \mnras 480(2):2597--2608, \doi{10.1093/mnras/sty1877},
  \eprint{1807.04529}

\bibitem[{{Rothman} and {Gordon}(2014)}]{rothman14}
{Rothman} LS, {Gordon} IE (2014) {Status of the HITRAN and HITEMP databases}.
  In: 13th International HITRAN Conference, p~49, \doi{10.5281/zenodo.11207}

\bibitem[{{Rothman} et~al.(2010){Rothman}, {Gordon}, {Barber}, {Dothe},
  {Gamache}, {Goldman}, {Perevalov}, {Tashkun}, and
  {Tennyson}}]{RothmanEtal2010jqsrtHITEMP}
{Rothman} LS, {Gordon} IE, {Barber} RJ, {Dothe} H, {Gamache} RR, {Goldman} A,
  {Perevalov} VI, {Tashkun} SA, {Tennyson} J (2010) {HITEMP, the
  high-temperature molecular spectroscopic database}. Journal of Quantitative
  Spectroscopy and Radiative Transfer 111:2139--2150, \doi{10.1016/j.jqsr
  t.2010.05.001}

\bibitem[{{Rothman} et~al.(2013){Rothman}, {Gordon}, {Babikov}, {Barbe}, {Chris
  Benner}, {Bernath}, {Birk}, {Bizzocchi}, {Boudon}, {Brown}, {Campargue},
  {Chance}, {Cohen}, {Coudert}, {Devi}, {Drouin}, {Fayt}, {Flaud}, {Gamache},
  {Harrison}, {Hartmann}, {Hill}, {Hodges}, {Jacquemart}, {Jolly}, {Lamouroux},
  {Le Roy}, {Li}, {Long}, {Lyulin}, {Mackie}, {Massie}, {Mikhailenko},
  {M{\"u}ller}, {Naumenko}, {Nikitin}, {Orphal}, {Perevalov}, {Perrin},
  {Polovtseva}, {Richard}, {Smith}, {Starikova}, {Sung}, {Tashkun}, {Tennyson},
  {Toon}, {Tyuterev}, and {Wagner}}]{Rothman2013}
{Rothman} LS, {Gordon} IE, {Babikov} Y, {Barbe} A, {Chris Benner} D, {Bernath}
  PF, {Birk} M, {Bizzocchi} L, {Boudon} V, {Brown} LR, {Campargue} A, {Chance}
  K, {Cohen} EA, {Coudert} LH, {Devi} VM, {Drouin} BJ, {Fayt} A, {Flaud} JM,
  {Gamache} RR, {Harrison} JJ, {Hartmann} JM, {Hill} C, {Hodges} JT,
  {Jacquemart} D, {Jolly} A, {Lamouroux} J, {Le Roy} RJ, {Li} G, {Long} DA,
  {Lyulin} OM, {Mackie} CJ, {Massie} ST, {Mikhailenko} S, {M{\"u}ller} HSP,
  {Naumenko} OV, {Nikitin} AV, {Orphal} J, {Perevalov} V, {Perrin} A,
  {Polovtseva} ER, {Richard} C, {Smith} MAH, {Starikova} E, {Sung} K, {Tashkun}
  S, {Tennyson} J, {Toon} GC, {Tyuterev} VG, {Wagner} G (2013) {The HITRAN2012
  molecular spectroscopic database}. \jqsrt 130:4--50,
  \doi{10.1016/j.jqsrt.2013.07.002}

\bibitem[{{Ryabchikova} et~al.(2015){Ryabchikova}, {Piskunov}, {Kurucz},
  {Stempels}, {Heiter}, {Pakhomov}, and {Barklem}}]{ryabchikova2015}
{Ryabchikova} T, {Piskunov} N, {Kurucz} RL, {Stempels} HC, {Heiter} U,
  {Pakhomov} Y, {Barklem} PS (2015) {A major upgrade of the VALD database}.
  Physica Scripta 90(5):054005, \doi{10.1088/0031-8949/90/5/054005}

\bibitem[{{Sing} et~al.(2016){Sing}, {Fortney}, {Nikolov}, {Wakeford},
  {Kataria}, {Evans}, {Aigrain}, {Ballester}, {Burrows}, {Deming},
  {D{\'e}sert}, {Gibson}, {Henry}, {Huitson}, {Knutson}, {Lecavelier Des
  Etangs}, {Pont}, {Showman}, {Vidal-Madjar}, {Williamson}, and
  {Wilson}}]{sing16}
{Sing} DK, {Fortney} JJ, {Nikolov} N, {Wakeford} HR, {Kataria} T, {Evans} TM,
  {Aigrain} S, {Ballester} GE, {Burrows} AS, {Deming} D, {D{\'e}sert} JM,
  {Gibson} NP, {Henry} GW, {Huitson} CM, {Knutson} HA, {Lecavelier Des Etangs}
  A, {Pont} F, {Showman} AP, {Vidal-Madjar} A, {Williamson} MH, {Wilson} PA
  (2016) {A continuum from clear to cloudy hot-Jupiter exoplanets without
  primordial water depletion}. \nat 529:59--62, \doi{10.1038/nature16068},
  \eprint{1512.04341}

\bibitem[{Skaf et~al.(2020)Skaf, Bieger, Edwards, Changeat, Morvan, Kiefer,
  Blain, Zingales, Poveda, Al-Refaie, Baeyens, Gressier, Guilluy, Jaziri,
  Modirrousta-Galian, Mugnai, Pluriel, Whiteford, Wright, Yip, Charnay,
  Leconte, Drossart, Tsiaras, Venot, Waldmann, and Beaulieu}]{skaf2020}
Skaf N, Bieger MF, Edwards B, Changeat Q, Morvan M, Kiefer F, Blain D, Zingales
  T, Poveda M, Al-Refaie A, Baeyens R, Gressier A, Guilluy G, Jaziri AY,
  Modirrousta-Galian D, Mugnai LV, Pluriel W, Whiteford N, Wright S, Yip KH,
  Charnay B, Leconte J, Drossart P, Tsiaras A, Venot O, Waldmann I, Beaulieu JP
  (2020) Ares ii: Characterising the hot jupiters wasp-127 b, wasp-79 b and
  wasp-62 b with hst. \eprint{2005.09615}

\bibitem[{{Stevenson} et~al.(2014){Stevenson}, {D{\'e}sert}, {Line}, {Bean},
  {Fortney}, {Showman}, {Kataria}, {Kreidberg}, {McCullough}, {Henry},
  {Charbonneau}, {Burrows}, {Seager}, {Madhusudhan}, {Williamson}, and
  {Homeier}}]{stevenson14b}
{Stevenson} KB, {D{\'e}sert} JM, {Line} MR, {Bean} JL, {Fortney} JJ, {Showman}
  AP, {Kataria} T, {Kreidberg} L, {McCullough} PR, {Henry} GW, {Charbonneau} D,
  {Burrows} A, {Seager} S, {Madhusudhan} N, {Williamson} MH, {Homeier} D (2014)
  {Thermal structure of an exoplanet atmosphere from phase-resolved emission
  spectroscopy}. Science 346:838--841, \doi{10.1126/science.1256758},
  \eprint{1410.2241}

\bibitem[{{Tennyson} and {Yurchenko}(2018)}]{TennysonYurchenko2018atomsExomol}
{Tennyson} J, {Yurchenko} S (2018) {The ExoMol Atlas of Molecular Opacities}.
  Atoms 6(2):26, \doi{10.3390/atoms6020026}, \eprint{1805.03711}

\bibitem[{{Tennyson} et~al.(2016){Tennyson}, {Yurchenko}, {Al-Refaie},
  {Barton}, {Chubb}, {Coles}, {Diamantopoulou}, {Gorman}, {Hill}, {Lam},
  {Lodi}, {McKemmish}, {Na}, {Owens}, {Polyansky}, {Rivlin}, {Sousa-Silva},
  {Underwood}, {Yachmenev}, and {Zak}}]{tennyson2016}
{Tennyson} J, {Yurchenko} SN, {Al-Refaie} AF, {Barton} EJ, {Chubb} KL, {Coles}
  PA, {Diamantopoulou} S, {Gorman} MN, {Hill} C, {Lam} AZ, {Lodi} L,
  {McKemmish} LK, {Na} Y, {Owens} A, {Polyansky} OL, {Rivlin} T, {Sousa-Silva}
  C, {Underwood} DS, {Yachmenev} A, {Zak} E (2016) {The ExoMol database:
  Molecular line lists for exoplanet and other hot atmospheres}. Journal of
  Molecular Spectroscopy 327:73--94, \doi{10.1016/j.jms.2016.05.002},
  \eprint{1603.05890}

\bibitem[{{Tinetti} et~al.(2016){Tinetti}, {Drossart}, {Eccleston}, {Hartogh},
  {Heske}, {Leconte}, {Micela}, {Ollivier}, {Pilbratt}, {Puig}, {Turrini},
  {Vand enbussche}, {Wolkenberg}, {Pascale}, {Beaulieu}, {G{\"u}del}, {Min},
  {Rataj}, {Ray}, {Ribas}, {Barstow}, {Bowles}, {Coustenis}, {Coud{\'e} du
  Foresto}, {Decin}, {Encrenaz}, {Forget}, {Friswell}, {Griffin}, {Lagage},
  {Malaguti}, {Moneti}, {Morales}, {Pace}, {Rocchetto}, {Sarkar}, {Selsis},
  {Taylor}, {Tennyson}, {Venot}, {Waldmann}, {Wright}, {Zingales}, and
  {Zapatero-Osorio}}]{tinetti16}
{Tinetti} G, {Drossart} P, {Eccleston} P, {Hartogh} P, {Heske} A, {Leconte} J,
  {Micela} G, {Ollivier} M, {Pilbratt} G, {Puig} L, {Turrini} D, {Vand
  enbussche} B, {Wolkenberg} P, {Pascale} E, {Beaulieu} JP, {G{\"u}del} M,
  {Min} M, {Rataj} M, {Ray} T, {Ribas} I, {Barstow} J, {Bowles} N, {Coustenis}
  A, {Coud{\'e} du Foresto} V, {Decin} L, {Encrenaz} T, {Forget} F, {Friswell}
  M, {Griffin} M, {Lagage} PO, {Malaguti} P, {Moneti} A, {Morales} JC, {Pace}
  E, {Rocchetto} M, {Sarkar} S, {Selsis} F, {Taylor} W, {Tennyson} J, {Venot}
  O, {Waldmann} IP, {Wright} G, {Zingales} T, {Zapatero-Osorio} MR (2016) {The
  science of ARIEL (Atmospheric Remote-sensing Infrared Exoplanet
  Large-survey)}. In: \procspie, Society of Photo-Optical Instrumentation
  Engineers (SPIE) Conference Series, vol 9904, p 99041X,
  \doi{10.1117/12.2232370}

\bibitem[{{Tinetti} et~al.(2018){Tinetti}, {Drossart}, {Eccleston}, {Hartogh},
  {Heske}, {Leconte}, {Micela}, {Ollivier}, {Pilbratt}, {Puig}, {Turrini},
  {Vandenbussche}, {Wolkenberg}, {Beaulieu}, {Buchave}, {Ferus}, {Griffin},
  {Guedel}, {Justtanont}, {Lagage}, {Machado}, {Malaguti}, {Min},
  {N{\o}rgaard-Nielsen}, {Rataj}, {Ray}, {Ribas}, {Swain}, {Szabo}, {Werner},
  {Barstow}, {Burleigh}, {Cho}, {du Foresto}, {Coustenis}, {Decin}, {Encrenaz},
  {Galand }, {Gillon}, {Helled}, {Morales}, {Mu{\~n}oz}, {Moneti}, {Pagano},
  {Pascale}, {Piccioni}, {Pinfield}, {Sarkar}, {Selsis}, {Tennyson}, {Triaud},
  {Venot}, {Waldmann}, {Waltham}, {Wright}, {Amiaux}, {Augu{\`e}res},
  {Berth{\'e}}, {Bezawada}, {Bishop}, {Bowles}, {Coffey}, {Colom{\'e}},
  {Crook}, {Crouzet}, {Da Peppo}, {Sanz}, {Focardi}, {Frericks}, {Hunt},
  {Kohley}, {Middleton}, {Morgante}, {Ottensamer}, {Pace}, {Pearson},
  {Stamper}, {Symonds}, {Rengel}, {Renotte}, {Ade}, {Affer}, {Alard}, {Allard},
  {Altieri}, {Andr{\'e}}, {Arena}, {Argyriou}, {Aylward}, {Baccani}, {Bakos},
  {Banaszkiewicz}, {Barlow}, {Batista}, {Bellucci}, {Benatti}, {Bernardi},
  {B{\'e}zard}, {Blecka}, {Bolmont}, {Bonfond}, {Bonito}, {Bonomo}, {Brucato},
  {Brun}, {Bryson}, {Bujwan}, {Casewell}, {Charnay}, {Pestellini}, {Chen},
  {Ciaravella}, {Claudi}, {Cl{\'e}dassou}, {Damasso}, {Damiano}, {Danielski},
  {Deroo}, {Di Giorgio}, {Dominik}, {Doublier}, {Doyle}, {Doyon}, {Drummond},
  {Duong}, {Eales}, {Edwards}, {Farina}, {Flaccomio}, {Fletcher}, {Forget},
  {Fossey}, {Fr{\"a}nz}, {Fujii}, {Garc{\'\i}a-Piquer}, {Gear}, {Geoffray},
  {G{\'e}rard}, {Gesa}, {Gomez}, {Graczyk}, {Griffith}, {Grodent}, {Guarcello},
  {Gustin}, {Hamano}, {Hargrave}, {Hello}, {Heng}, {Herrero}, {Hornstrup},
  {Hubert}, {Ida}, {Ikoma}, {Iro}, {Irwin}, {Jarchow}, {Jaubert}, {Jones},
  {Julien}, {Kameda}, {Kerschbaum}, {Kervella}, {Koskinen}, {Krijger}, {Krupp},
  {Lafarga}, {Landini}, {Lellouch}, {Leto}, {Luntzer}, {Rank-L{\"u}ftinger},
  {Maggio}, {Maldonado}, {Maillard}, {Mall}, {Marquette}, {Mathis}, {Maxted},
  {Matsuo}, {Medvedev}, {Miguel}, {Minier}, {Morello}, {Mura}, {Narita},
  {Nascimbeni}, {Nguyen Tong}, {Noce}, {Oliva}, {Palle}, {Palmer}, {Pancrazzi},
  {Papageorgiou}, {Parmentier}, {Perger}, {Petralia}, {Pezzuto},
  {Pierrehumbert}, {Pillitteri}, {Piotto}, {Pisano}, {Prisinzano}, {Radioti},
  {R{\'e}ess}, {Rezac}, {Rocchetto}, {Rosich}, {Sanna}, {Santerne}, {Savini},
  {Scandariato}, {Sicardy}, {Sierra}, {Sindoni}, {Skup}, {Snellen}, {Sobiecki},
  {Soret}, {Sozzetti}, {Stiepen}, {Strugarek}, {Taylor}, {Taylor}, {Terenzi},
  {Tessenyi}, {Tsiaras}, {Tucker}, {Valencia}, {Vasisht}, {Vazan}, {Vilardell},
  {Vinatier}, {Viti}, {Waters}, {Wawer}, {Wawrzaszek}, {Whitworth}, {Yung},
  {Yurchenko}, {Osorio}, {Zellem}, {Zingales}, and {Zwart}}]{tinetti18}
{Tinetti} G, {Drossart} P, {Eccleston} P, {Hartogh} P, {Heske} A, {Leconte} J,
  {Micela} G, {Ollivier} M, {Pilbratt} G, {Puig} L, {Turrini} D,
  {Vandenbussche} B, {Wolkenberg} P, {Beaulieu} JP, {Buchave} LA, {Ferus} M,
  {Griffin} M, {Guedel} M, {Justtanont} K, {Lagage} PO, {Machado} P, {Malaguti}
  G, {Min} M, {N{\o}rgaard-Nielsen} HU, {Rataj} M, {Ray} T, {Ribas} I, {Swain}
  M, {Szabo} R, {Werner} S, {Barstow} J, {Burleigh} M, {Cho} J, {du Foresto}
  VC, {Coustenis} A, {Decin} L, {Encrenaz} T, {Galand } M, {Gillon} M, {Helled}
  R, {Morales} JC, {Mu{\~n}oz} AG, {Moneti} A, {Pagano} I, {Pascale} E,
  {Piccioni} G, {Pinfield} D, {Sarkar} S, {Selsis} F, {Tennyson} J, {Triaud} A,
  {Venot} O, {Waldmann} I, {Waltham} D, {Wright} G, {Amiaux} J, {Augu{\`e}res}
  JL, {Berth{\'e}} M, {Bezawada} N, {Bishop} G, {Bowles} N, {Coffey} D,
  {Colom{\'e}} J, {Crook} M, {Crouzet} PE, {Da Peppo} V, {Sanz} IE, {Focardi}
  M, {Frericks} M, {Hunt} T, {Kohley} R, {Middleton} K, {Morgante} G,
  {Ottensamer} R, {Pace} E, {Pearson} C, {Stamper} R, {Symonds} K, {Rengel} M,
  {Renotte} E, {Ade} P, {Affer} L, {Alard} C, {Allard} N, {Altieri} F,
  {Andr{\'e}} Y, {Arena} C, {Argyriou} I, {Aylward} A, {Baccani} C, {Bakos} G,
  {Banaszkiewicz} M, {Barlow} M, {Batista} V, {Bellucci} G, {Benatti} S,
  {Bernardi} P, {B{\'e}zard} B, {Blecka} M, {Bolmont} E, {Bonfond} B, {Bonito}
  R, {Bonomo} AS, {Brucato} JR, {Brun} AS, {Bryson} I, {Bujwan} W, {Casewell}
  S, {Charnay} B, {Pestellini} CC, {Chen} G, {Ciaravella} A, {Claudi} R,
  {Cl{\'e}dassou} R, {Damasso} M, {Damiano} M, {Danielski} C, {Deroo} P, {Di
  Giorgio} AM, {Dominik} C, {Doublier} V, {Doyle} S, {Doyon} R, {Drummond} B,
  {Duong} B, {Eales} S, {Edwards} B, {Farina} M, {Flaccomio} E, {Fletcher} L,
  {Forget} F, {Fossey} S, {Fr{\"a}nz} M, {Fujii} Y, {Garc{\'\i}a-Piquer} {\'A},
  {Gear} W, {Geoffray} H, {G{\'e}rard} JC, {Gesa} L, {Gomez} H, {Graczyk} R,
  {Griffith} C, {Grodent} D, {Guarcello} MG, {Gustin} J, {Hamano} K, {Hargrave}
  P, {Hello} Y, {Heng} K, {Herrero} E, {Hornstrup} A, {Hubert} B, {Ida} S,
  {Ikoma} M, {Iro} N, {Irwin} P, {Jarchow} C, {Jaubert} J, {Jones} H, {Julien}
  Q, {Kameda} S, {Kerschbaum} F, {Kervella} P, {Koskinen} T, {Krijger} M,
  {Krupp} N, {Lafarga} M, {Landini} F, {Lellouch} E, {Leto} G, {Luntzer} A,
  {Rank-L{\"u}ftinger} T, {Maggio} A, {Maldonado} J, {Maillard} JP, {Mall} U,
  {Marquette} JB, {Mathis} S, {Maxted} P, {Matsuo} T, {Medvedev} A, {Miguel} Y,
  {Minier} V, {Morello} G, {Mura} A, {Narita} N, {Nascimbeni} V, {Nguyen Tong}
  N, {Noce} V, {Oliva} F, {Palle} E, {Palmer} P, {Pancrazzi} M, {Papageorgiou}
  A, {Parmentier} V, {Perger} M, {Petralia} A, {Pezzuto} S, {Pierrehumbert} R,
  {Pillitteri} I, {Piotto} G, {Pisano} G, {Prisinzano} L, {Radioti} A,
  {R{\'e}ess} JM, {Rezac} L, {Rocchetto} M, {Rosich} A, {Sanna} N, {Santerne}
  A, {Savini} G, {Scandariato} G, {Sicardy} B, {Sierra} C, {Sindoni} G, {Skup}
  K, {Snellen} I, {Sobiecki} M, {Soret} L, {Sozzetti} A, {Stiepen} A,
  {Strugarek} A, {Taylor} J, {Taylor} W, {Terenzi} L, {Tessenyi} M, {Tsiaras}
  A, {Tucker} C, {Valencia} D, {Vasisht} G, {Vazan} A, {Vilardell} F,
  {Vinatier} S, {Viti} S, {Waters} R, {Wawer} P, {Wawrzaszek} A, {Whitworth} A,
  {Yung} YL, {Yurchenko} SN, {Osorio} MRZ, {Zellem} R, {Zingales} T, {Zwart} F
  (2018) {A chemical survey of exoplanets with ARIEL}. Experimental Astronomy
  46(1):135--209, \doi{10.1007/s10686-018-9598-x}

\bibitem[{{Toon} and {Ackerman}(1981)}]{ToonAckerman1981apoptScattering}
{Toon} OB, {Ackerman} TP (1981) {Algorithms for the calculation of scattering
  by stratified spheres}. \ao 20(20):3657--3660, \doi{10.1364/AO.20.003657}

\bibitem[{{Tsang} et~al.(2010){Tsang}, {Wilson}, {Barstow}, {Irwin}, {Taylor},
  {McGouldrick}, {Piccioni}, {Drossart}, and {Svedhem}}]{tsang10}
{Tsang} CCC, {Wilson} CF, {Barstow} JK, {Irwin} PGJ, {Taylor} FW, {McGouldrick}
  K, {Piccioni} G, {Drossart} P, {Svedhem} H (2010) {Correlations between cloud
  thickness and sub-cloud water abundance on Venus}. \grl 37(2):L02202,
  \doi{10.1029/2009GL041770}

\bibitem[{{Tsiaras} et~al.(2018){Tsiaras}, {Waldmann}, {Zingales}, {Rocchetto},
  {Morello}, {Damiano}, {Karpouzas}, {Tinetti}, {McKemmish}, {Tennyson}, and
  {Yurchenko}}]{tsiaras18}
{Tsiaras} A, {Waldmann} IP, {Zingales} T, {Rocchetto} M, {Morello} G, {Damiano}
  M, {Karpouzas} K, {Tinetti} G, {McKemmish} LK, {Tennyson} J, {Yurchenko} SN
  (2018) {A Population Study of Gaseous Exoplanets}. \aj 155:156,
  \doi{10.3847/1538-3881/aaaf75}, \eprint{1704.05413}

\bibitem[{{Tsiaras} et~al.(2019){Tsiaras}, {Waldmann}, {Tinetti}, {Tennyson},
  and {Yurchenko}}]{tsiaras19}
{Tsiaras} A, {Waldmann} IP, {Tinetti} G, {Tennyson} J, {Yurchenko} SN (2019)
  {Water vapour in the atmosphere of the habitable-zone eight-Earth-mass planet
  K2-18 b}. {Nature Astronomy} 3:1086--1091, \doi{10.1038/s41550-019-0878-9},
  \eprint{1909.05218}

\bibitem[{{Venot} et~al.(2012){Venot}, {H{\'e}brard}, {Ag{\'u}ndez},
  {Dobrijevic}, {Selsis}, {Hersant}, {Iro}, and {Bounaceur}}]{venot01}
{Venot} O, {H{\'e}brard} E, {Ag{\'u}ndez} M, {Dobrijevic} M, {Selsis} F,
  {Hersant} F, {Iro} N, {Bounaceur} R (2012) {A chemical model for the
  atmosphere of hot Jupiters}. \aap 546:A43, \doi{10.1051/0004-6361/201219310},
  \eprint{1208.0560}

\bibitem[{{Waldmann} et~al.(2015{\natexlab{a}}){Waldmann}, {Rocchetto},
  {Tinetti}, {Barton}, {Yurchenko}, and {Tennyson}}]{waldmann2015b}
{Waldmann} IP, {Rocchetto} M, {Tinetti} G, {Barton} EJ, {Yurchenko} SN,
  {Tennyson} J (2015{\natexlab{a}}) {Tau-REx II: Retrieval of Emission
  Spectra}. \apj 813(1):13, \doi{10.1088/0004-637X/813/1/13},
  \eprint{1508.07591}

\bibitem[{{Waldmann} et~al.(2015{\natexlab{b}}){Waldmann}, {Tinetti},
  {Rocchetto}, {Barton}, {Yurchenko}, and {Tennyson}}]{waldmann2015a}
{Waldmann} IP, {Tinetti} G, {Rocchetto} M, {Barton} EJ, {Yurchenko} SN,
  {Tennyson} J (2015{\natexlab{b}}) {Tau-REx I: A Next Generation Retrieval
  Code for Exoplanetary Atmospheres}. \apj 802:107,
  \doi{10.1088/0004-637X/802/2/107}, \eprint{1409.2312}

\bibitem[{{Yurchenko} et~al.(2017){Yurchenko}, {Amundsen}, {Tennyson}, and
  {Waldmann}}]{yurchenko17}
{Yurchenko} SN, {Amundsen} DS, {Tennyson} J, {Waldmann} IP (2017) {A hybrid
  line list for CH$_{4}$ and hot methane continuum}. \aap 605:A95,
  \doi{10.1051/0004-6361/201731026}, \eprint{1706.05724}

\bibitem[{{Zhang} et~al.(2019){Zhang}, {Chachan}, {Kempton}, and
  {Knutson}}]{zhang2019}
{Zhang} M, {Chachan} Y, {Kempton} EMR, {Knutson} HA (2019) {Forward Modeling
  and Retrievals with PLATON, a Fast Open-source Tool}. Publications of the
  Astronomical Society of the Pacific 131(997):034501,
  \doi{10.1088/1538-3873/aaf5ad}, \eprint{1811.11761}

\end{thebibliography}


%
%

\end{document}